\documentclass[letter,11pt]{article}

\usepackage{jheppub}
\usepackage{hyperref}
\usepackage{graphicx}
\usepackage{amsmath}
\usepackage{amssymb}
\usepackage{xspace}
\usepackage{slashed}
\usepackage{subcaption}
\usepackage[normalem]{ulem}
\usepackage[dvipsnames]{xcolor}
\usepackage[utf8]{inputenc}
\usepackage[T1]{fontenc}
\usepackage{mathtools}
\usepackage{stmaryrd}
\usepackage{enumerate}
\usepackage{mathrsfs}
\usepackage[export]{adjustbox}

\def\veps{\varepsilon}
\def\eps{\epsilon}
\newcommand{\ra}{\rightarrow}
\newcommand{\lra}{\leftrightarrow}
\newcommand{\bn}{{\bar{n}}}

\newcommand{\nslash}{n\!\!\!\slash}
\newcommand{\bnslash}{\bar{n}\!\!\!\slash}

\def\df{{\rm d}}
\def\dfbar{{\mathchar'26\mkern-12mu d}}
\def\dfbar{{\mathchar'26\mkern-12mu {\rm d}}}

\newcommand{\nn}{\nonumber}
\def\nn{\nonumber}

\newcommand{\mb}[1]{\mathbf{#1}}
\newcommand{\im}{\mathrm{i}}

\newcommand{\eftn}{\ensuremath{{\rm EFT}_{n}}\xspace}
\newcommand{\eftnp}{\ensuremath{{\rm EFT}_{n+1}}\xspace}

\renewcommand{\Im}{\mathrm{Im}}
\renewcommand{\Re}{\mathrm{Re}}

\newcommand{\bl}[1]{\textcolor{Blue}{#1}}
\newcommand{\gr}[1]{\textcolor{Green}{#1}}

\newcommand{\eq}[1]{Eq.~\eqref{eq:#1}}
\newcommand{\eqs}[2]{Eqs.~\eqref{eq:#1} and \eqref{eq:#2}}
\renewcommand{\sec}[1]{Sec.~\ref{sec:#1}}

\newcommand{\secs}[2]{Secs.~\ref{sec:#1} and \ref{sec:#2}}

\newcommand{\app}[1]{App.~\ref{app:#1}}
\newcommand{\fig}[1]{Fig.~\ref{fig:#1}}
\newcommand{\figs}[2]{Figs.~\ref{fig:#1} and \ref{fig:#2}}

\DeclareRobustCommand{\Ref}[1]{Ref.~\cite{#1}}
\DeclareRobustCommand{\Refs}[1]{Refs.~\cite{#1}}

\DeclareRobustCommand{\eq}[1]{Eq.~(\ref{eq:#1})}
\DeclareRobustCommand{\eqs}[2]{Eqs.~(\ref{eq:#1}) and (\ref{eq:#2})}

\usepackage{marginnote}

\title{A new form of QCD coherence for multiple soft emissions using Glauber-SCET}

\author[a]{Aditya Pathak}
\affiliation[a]{University of Manchester, School of Physics and Astronomy, Manchester, M13 9PL, United Kingdom}

\abstract{Amplitude-level factorization for a soft gluon emission has long been understood in terms of a product of loop-expanded soft-gluon currents and hard scattering matrix elements, both of which are infrared (IR) divergent. Thus, the amplitude for multiple soft gluon emissions, ordered in their relative softness, can be written as a product of IR divergent soft gluon currents and the matrix elements.
In a more recent work, Angeles-Martinez, Forshaw and Seymour~\cite{Angeles-Martinez:2016dph} (AMFS) showed that the result for this amplitude can in fact be re-expressed in an ordered evolution approach, involving \textit{IR finite} one-loop insertions where the virtual loop momentum is constrained in a highly non-trivial way by the $k_T$ of the adjacent real emissions. The result thus exhibits a novel amplitude level QCD coherence where the IR divergences originating only from the very last, softest, gluon emission remain, and the rest cancel.
The proof of the AMFS result at one-loop in QCD, however, involves many diagrams, and only after carefully grouping and summing over all the diagrams does the correct ordering variable emerge, making the higher order extension a challenging task. Moreover, the compact, Markovian nature of the final AMFS result is suggestive of a deeper underlying physics that is obscured in the derivation using traditional diagrammatic QCD.
By considering a (recursive) sequence of effective field theories (EFTs) with Glauber-SCET operators, we present an elegant derivation of this result involving only a handful of diagrams.
The SCET derivation offers clean physical insights, and makes a higher order extension of the AMFS result tractable.
We also show that the grouping of QCD graphs necessary to derive the AMFS result in full theory is already implicit in the Feynman rules of Glauber-SCET operators such that the same result can alternatively be derived with significantly less effort in a single EFT with multiple ordered soft gluon emissions.}


\begin{document}
\maketitle

\section{Introduction}
\label{sec:Intro}
With lack of clear signatures of new physics in the high-energy data at the LHC, importance of precision Standard Model measurements cannot be overemphasized. Of central importance in the high-energy analyses are the general purpose parton showers that describe the evolution of a high energy parton as it radiates and makes its way to the detectors. The accuracy of existing parton showers at the cross section level, however, is at most next-to-leading-logarithmic and leading-color. Furthermore, for observables that entail a non-global measurement~\cite{Dasgupta:2001sh}, such as veto on emissions in the region (gap) between jets (or the beam region)~\cite{Oderda:1998en}, these parton showers fail to provide adequate description. Here the observable receives logarithmically enhanced contributions from arbitrary number of emissions in the jet and the beam (out of the gap) region. Accordingly, to capture the delicate quantum interference in such observables one needs to track the evolution of partons at the amplitude level~\cite{Nagy:2017ggp,Forshaw:2020wrq,Forshaw:2019ver}.

Factorization of soft and collinear contributions, which is the basis for parton showers and analytical approaches, is, however, not a universal property of QCD matrix elements, and is violated~\cite{Catani:2011st} by exchanges of offshell gluons that are instantaneous in the directions perpendicular to the momenta of partons between which they are exchanged~\cite{Lipatov:1996ts}. These gluons are said to obey ``Coulomb/Glauber scaling'' and lead to an imaginary contribution to the amplitude. In case of QED, they lead to an irrelevant phase.
However, for hadron collisions
they give rise to well known uncanceled ``superleading logarithms''~\cite{Forshaw:2006fk,Forshaw:2012bi,Becher:2021zkk}, that, although they do not appear for the first few orders in perturbation theory, are formally of higher logarithmic enhancement than the leading (single) logarithms.
Resummation of the aforementioned non-global logarithms (NGLs) at subleading color accuracy can be carried out in the algorithmic framework outlined in \Ref{Martinez:2018ffw}.
The algorithm involves considering a chain of emissions ordered in a kinematic variable, such as the energy or the transverse momentum, and interleaving the emission operators by insertions of Sudakov exponentials that account for virtual graphs, but with the limits of these insertions bounded by the order parameter of the adjacent emissions, and hence are IR finite. For processes that are insensitive to Glauber gluon exchanges the choice of ordering parameter is irrelevant, and the algorithm can be shown to obey an evolution equation that establishes equivalence with other approaches~\cite{Weigert:2003mm,Caron-Huot:2015bja,Larkoski:2015zka,Becher:2016mmh,Banfi:2021owj}. However, for Glauber-sensitive processes, such as the diagrams that lead to superleading logs, the choice of the ordering variable becomes a delicate issue~\cite{Banfi:2010xy}.

With the motivation of pinning down the nature of the ordering variable in presence of Glauber exchanges, the authors (AMFS) of \Refs{Angeles-Martinez:2015rna,Angeles-Martinez:2016dph} considered the amplitude for emission of multiple soft gluon emissions ordered in softness. They showed that at one-loop accuracy the amplitude for multiple soft gluon emissions (\eq{AMFS} below), that are ordered in their relative softness, can be expressed in a form analogous to the algorithm for NGLs presented in \Ref{Martinez:2018ffw}. However, the key difference here is that one does not \textit{pick} an ordering variable, but instead, the exact one-loop computation makes explicit
that transverse momenta of soft gluons evaluated in an appropriate dipole frame of adjecent real emissions bound the virtual loop integrals and render them finite. As a result, one sees a remarkable coherent cancellation of all the intermediate IR divergences between real and virtual graphs. Only the IR divergence from the very last soft gluon emission in the chain remain uncanceled.
The emergence of dipole transverse momenta that render the loop integrals finite can then be interpreted as an ordering parameter singled out by QCD.
Additionally, the specific choice of the dipole frame imposed upon us by this result prohibits a straightforward exponentiation of soft gluon emissions, and does not satisfy an evolution equation. This is another manifestation of how Glauber gluons generally destroy coherence.
The result, however, still retains the Markovian nature, which makes it amenable to development of future all-orders amplitude level parton shower. It is then interesting to see how the result generalizes to higher orders; see, for example, \Ref{Platzer:2020lbr} for efforts towards description of color flow evolution at two loops, which can shed interesting lights on the extension of AMFS result to higher orders.

The derivation of the AMFS result, as presented in \Ref{Angeles-Martinez:2015rna} using full theory (QCD) diagrams, however, follows only after careful grouping and summing over many QCD diagrams, with the intermediate steps bearing little resemblance with the final expression. Thus, an extension of this result to higher orders using a direct diagrammatic approach is an extremely challenging task. Fortunately, the effective field theory framework can help us here.
In this work we rederive this expression in the framework of soft collinear effective theory (SCET)~\cite{Bauer:2000ew,Bauer:2000yr,Bauer:2001yt,Bauer:2001ct,Bauer:2002nz} with Glauber potential operators~\cite{Rothstein:2016bsq}. We consider a recursive sequence of EFTs associated with each soft emission, where in passing from one EFT to another the corresponding soft emission is frozen to become a collinear mode in the next EFT, with fluctuations in the virtuality further restricted. This is achieved by appropriately modifying the hard scattering operator and evaluating the corresponding Wilson coefficient. The result then follows straightforwardly by considering combination of the Wilson coefficient and the matrix element for a single soft gluon emission in the low energy EFT.

We also verify the results by working in a single EFT and considering the ordered limit of the double soft gluon emission amplitude. This is analogous to the approach taken in \Ref{Angeles-Martinez:2015rna}, but here we make use of SCET diagrams instead of full theory graphs.
We will find that the calculation in SCET is organized in such a way that the necessary grouping of the QCD diagrams is already implicit in the EFT diagrams, which drastically simplifies the analysis involving a lot fewer diagrams. While SCET is most widely employed for facilitating higher order resummation, its application here to enable efficient fixed order computation is somewhat novel.
We will see a special role played by the Lipatov vertex (and its multiple gluon-generalizations) to cancel the intermediate IR divergences and make the coherent property of the result explicit. The derivation also makes it clear that the two-loop generalization of the AMFS result will involve the one-loop effective Lipatov vertex, the collinear-Glauber and the soft-Glauber vertices. We leave this analysis to future work.

The outline of the paper is as follows: in \sec{AMFS} we introduce and describe the AMFS result. The results for one-gluon emissions are derived in \sec{OneGluon}. Here we also describe the notation and the EFT setup. In \sec{twoGluon} we derive the results for two ordered, soft gluon emission amplitude and conclude the derivation of the AMFS result. The calculations in \secs{OneGluon}{twoGluon} provide the details for the results presented in the companion paper, \Ref{Forshaw:2021xxx}.
In \sec{DoubleSoft} we evaluate the amplitude for double soft emissions in SCET and consider the limit of one gluon being yet softer than the other, and verify the results of the previous section and make a connection with the derivation presented in \Ref{Angeles-Martinez:2015rna}. We conclude and discuss future directions in \sec{conclusion}. The discussion of the color-space notation, Feynman rules, and explicit calculations of individual diagrams are delegated to the appendices.
\section{The AMFS result}
\label{sec:AMFS}
Here we set up the notation and review the AMFS result for amplitude of multiple ordered soft gluon emissions. We consider a process involving $n$ hard partons and additional soft gluon amplitudes. The amplitude for $n$ hard partons can be written as
\begin{align}
{\cal M}^{\{a\}}(\{p_i\}) = \big[ \{a\}\big | {\cal M} (p_1,\ldots , p_n)\big] \, ,
\end{align}
where $\{p_i\}$ denote the momenta and $\{a\}$ are the color indices of the $n$ partons, and $\big| {\cal M}\big]$ is a vector in color space. At leading power, amplitude for an additional soft emission factorizes~\cite{Bassetto:1983mvz,Bern:1999ry,CATANI2000435,Duhr:2013msa,Li:2013lsa,Feige:2014wja}
\begin{align}\label{eq:Ma1}
{\cal M}^{C\cup\{a\}}(q, \veps;\{p_i\}) = \big[ C\cup \{a\}\big | {\cal M} ({q}, p_1,\ldots p_n) \big ] &\simeq g \mu^\eps
\veps^*_\mu ({q}) \big[ C\cup \{a\}\big | \mb J^\mu ({q}) \big | {\cal M} (p_1,\ldots , p_n) \big] \, ,
\end{align}
where $ q$ is the soft gluon momentum, $\veps(q)$ the polarization vector, and $C$ is the adjoint color index.
The factorization in \eq{JMFact1} holds in the limit where $q \sim \lambda p_i$, $\lambda \ll1$.

Next, both the hard matrix element and soft gluon amplitudes have the loop expansion:
\begin{align}\label{eq:JMFact1}
| {\cal M} (p_1,\ldots , p_n) ] &= \big| M_0^{(0)}\big ] + \big| M_0^{(1)}\big ] + \ldots
\, , \qquad
\mb J({q}) = \mb J^{(0)} (q) + \mb J^{(1)} (q) + \ldots \, ,
\end{align}
where the superscripts ${(0,1)}$ denote the order of loop-expansion of the soft gluon current and the hard matrix element, both of which are IR divergent. The tree level soft current is given by
\begin{align}\label{eq:J0}
\mb J^{(0)\mu} (q) &= \sum_{i = 1}^{n} \mb T_i \frac{p_i^\mu}{p_i\cdot q}
= \sum_{j = 1}^{n} \mb d^{(0)\mu}_{ij}(q) \, , \qquad
\mb d^{(0)\mu}_{ij}(q) = \mb T_j \Big(\frac{p_j^\mu}{ p_j \cdot q} - \frac{p_i^\mu}{p_i \cdot q}\Big)
\, ,
\end{align}
where the tree level operator $\mb d^{(0)\mu}_{ij}(q)$ describes coherent emission of the soft gluon off the hard partons $i$ and $j$. Note that due to color conservation the choice of the index $j$ on $\mb d^{(0)\mu}_{ij}$ is completely arbitrary. Here, $\mb J$ and $\mb d_{ij}$ without the Lorentz index $\mu$ include polarization vectors. For example,
\begin{align}
\mb J(q) \equiv \veps_\mu^*(q) \mb J^\mu(q) \, , \qquad \mb d_{ij}^{(0)} (q) \equiv \veps_\mu^*(q) \mb d^{(0)\mu}_{ij}(q) \, .
\end{align}

\Ref{Angeles-Martinez:2016dph} considered the matrix element for $ N$ soft gluon emissions with momenta $\{q_{m}\}$ ordered in their relative softness i.e. $q_{m+1} \sim \lambda q_{m}$. Using the \eq{JMFact1} the result can be expressed as a product of soft current, where each soft gluon acts as a source for the subsequent, yet softer emission. Thus, the final expression obtained can be written as
\begin{align}\label{eq:JMFactN}
\big | M_{{N}} \big ] &= (g \mu^\eps)^{{N}} \mb J ({q_N}) \ldots \mb J ({q_1})
\big| {\cal M} (p_1 , \ldots, p_n) \big ] \, ,
\end{align}
At one-loop accuracy, we have
\begin{align} \label{eq:OneLoop}
\mb J^{(1)} ({q_{m+1}}) &= \frac{1}{2} \sum_{j = 1}^{n+{m}}\sum_{k=1}^{n+{m}} \mb d_{jk}^{(1)}({q_{m+1}}) \,, \quad
\big| M_0^{(1)}\big ] = \sum_{i = 2}^{n} \sum_{j=1}^{i -1} \mb I^{(ij)}(0, \omega_{ij}) \big | M_0^{(0)} \big ] \, ,
\quad
\omega_{ij} \equiv 2p_i \cdot p_j \, ,
\end{align}
where
\begin{align}\label{eq:d1ijDef}
\mb d_{ij}^{(1)} (q) &\equiv \frac{\alpha_s}{2\pi} \frac{c_\Gamma}{\eps^2} \mb T_{q} \cdot \mb T_i \Bigg(\frac{ e^{-\im \pi \tilde \delta_{ij}}}{e^{-\im \pi \tilde \delta_{iq}}e^{-\im \pi \tilde \delta_{jq}}} \frac{4\pi \mu^2}{\big({q_{\perp}^{(ij)}}\big)^2}\Bigg)^\eps
\mb d^{(0)}_{ij}(q) \, , \qquad c_\Gamma = \frac{\Gamma^3(1-\eps)\Gamma^2(1+\eps)}{\Gamma(1-2\eps)} \, ,
\nn \\
\mb I^{(ij)}(0,\omega_{ij}) &\equiv
\frac{\alpha_s}{2\pi}
\frac{ c_\Gamma}{\eps^2}
\mb T_i \cdot \mb T_j
\Bigg(e^{\im \pi \tilde \delta_{ij}}\frac{4\pi \mu^2 }{\omega_{ij}}\Bigg)^\eps
\, ,
\end{align}
where the insertions $\mb I^{(ij)}$'s correspond to virtual emissions exchanged between the hard legs $i$ and $j$. We have $\tilde \delta_{ij} = 1$ when both the emissions are incoming or outgoing, and zero otherwise.
The transverse momentum ${q_{\perp}^{(ij)}}$ is defined in the $(ij)$ dipole frame:
\begin{align}\label{eq:qperpijDef}
\big( {q_{\perp}^{(ij)} }\big)^2= \frac{2 p_i \cdot q \, p_j \cdot q}{p_i \cdot p_j} \, ,
\end{align}
In \eq{JMFactN} we see that every soft emission contributes to the set of sources for the subsequent soft emissions. The $0$ in the first argument in \eq{OneLoop} indicates that the insertion is IR divergent, regulated by $\eps$. Thus, all the one-loop factors in \eq{JMFactN} are IR divergent, and it is not clear if there are any intermediate cancellations upon expanding the $\mb d^{(1)}_{ij}$ and $\mb I^{(ij)}$ insertions.

It was shown in \Ref{Angeles-Martinez:2016dph} that the result in \eq{JMFactN} can be equivalently expressed in an ``ordered evolution approach'':
\begin{align}\label{eq:AMFS}
&\big | M^{(1)}_{\gr N} \big ] =
(g\mu^{\eps})^{\textcolor{Green}{N}} \bigg( \prod_{k = 1}^{\gr N} \mb J^{(0)}(\textcolor{Green}{q_k})\bigg)
\bigg(
\sum_{\textcolor{Red}{i} = 2}^{n} \sum_{\textcolor{Red}{j} < \textcolor{Red}{i}}
\mathbf{I}^{(\textcolor{Red}{ij})}( \textcolor{Green}{q_{1\perp}^{(\textcolor{Red}{ij})}}, \textcolor{Red}{\sqrt{\omega_{ij}}})
\bigg) \big|M_0^{(0)} \big ]
\\
&+ (g\mu^{\eps})^{\textcolor{Green}{N}} \sum_{m= 1}^{\textcolor{Green}{N}}
\bigg( \prod_{k = {m} + 1}^{\gr N} \mb J^{(0)}(\textcolor{Green}{q_k})\bigg)
\bigg(
\sum_{\textcolor{Red}{i} = 2}^{n + m-1} \sum_{\textcolor{Red}{j} <\textcolor{Red}{i}}
\mb I^{(\textcolor{Red}{ij})} ( \textcolor{Green}{q_{m+1\perp}^{(\textcolor{Red}{ij})}},
\bl{q^{(\textcolor{Red}{ij})}_{m\perp}})\,
\bigg)
\bigg( \prod_{\ell = 1}^{m} \mb J^{(0)}(\bl{q_\ell})\bigg)
\big | M_0^{(0)} \big ]
\nn
\\
&+ (g\mu^{\eps})^{\textcolor{Green}{N}} \sum_{m = 1}^{\textcolor{Green}{N}}
\bigg( \prod_{k = {m} + 1}^{\gr N} \mb J^{(0)}(\textcolor{Green}{q_k})\bigg)
\bigg(
\sum_{\textcolor{Red}{i}, \textcolor{Red}{j} = 1}^{n+m-1}
\mathbf{I}^{(\textcolor{Blue}{(n+m) }\textcolor{Red}{i})} ( \textcolor{Green}{q_{m+1\perp}^{(\textcolor{Blue}{(n+m)}\textcolor{Red}{i})}},
\textcolor{Blue}{q_{m\perp}^{(\textcolor{Red}{ij})}})
\mb d^{(0)}_{\textcolor{Red}{ij}} (\textcolor{Blue}{q_m}) \bigg)
\bigg( \prod_{\ell = 1}^{m - 1} \mb J^{(0)}(\bl{q_\ell})\bigg)
\big | M^{(0)}_0 \big ]
\nn \, ,
\end{align}
where
\begin{align}
&\mb I^{(ij)}(a,b) = \frac{\alpha_s}{2\pi}\frac{1}{\eps^2} \mb T_i \cdot \mb T_j
\bigg[
\Big(\frac{4\pi \mu^2}{b^2}\Big)^\eps
\Big(
1 + \im \pi\eps \tilde \delta_{ij} - \eps \ln \frac{\omega_{ij}}{b^2}
\Big)
-
\Big(\frac{4\pi \mu^2}{a^2}\Big)^\eps
\Big(
1 + \im \pi \eps \tilde \delta_{ij} - \eps \ln \frac{\omega_{ij}}{a^2}
\Big)
\bigg] + {\rm Re}~ {\cal O}(\eps)\nn \\
&\quad= \frac{\alpha_s}{2\pi}\frac{1}{\eps^2} \mb T_i \cdot \mb T_j
\bigg[
\Big(\frac{4\pi \mu^2}{b^2}\Big)^\eps
\Big(
1 - \eps \ln \frac{-\omega_{ij}}{b^2}
\Big)
-
\Big(\frac{4\pi \mu^2}{a^2}\Big)^\eps
\Big(
1 - \eps \ln \frac{-\omega_{ij}}{a^2}
\Big)
\bigg] + {\rm Re}~ {\cal O}(\eps) \, ,
\end{align}
where the real part is correct up to non-logarithmic terms. We have rewritten the equation in the second line to emphasize that the real part can be obtained by analytically continuing the result for the imaginary part.
\begin{figure}[t!]
\centering
\includegraphics[width=\textwidth]{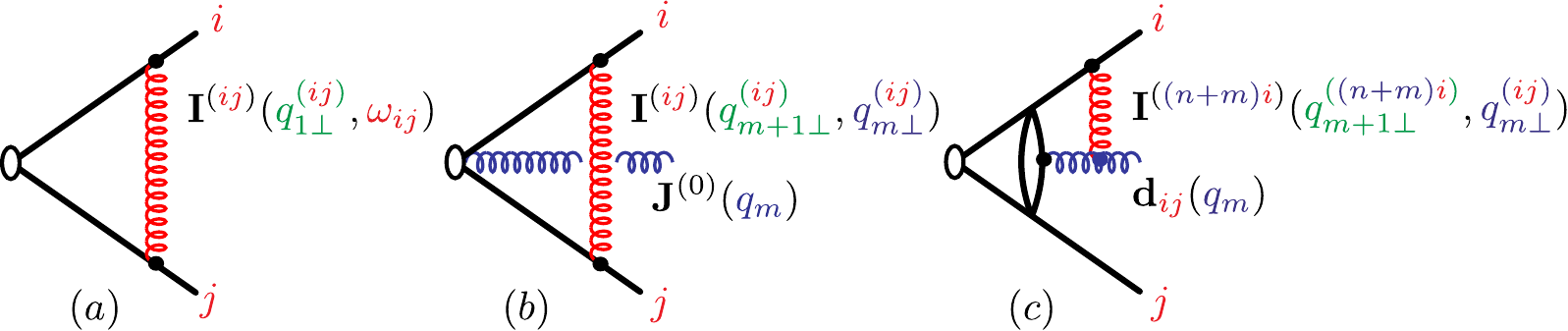}
\caption{Relation of limits of the virtual loop insertions to the adjacent dipole emission momenta in the ordered soft gluon emission result. The graphs (a), (b) and (c) respectively correspond to the three lines in \eq{AMFS}.}
\label{fig:AMFS}
\end{figure}

Unlike \eq{JMFactN}, the insertions $\mb I^{(ij)}$ are, however, IR finite with the loop momentum $k_\perp$ of the bounded above and below by its arguments due to non-trivial coherent cancellation of various contributions.
The result can be understood by relating each of the lines by the three diagrams shown in \fig{AMFS}.
The first line involves tree level soft gluon currents, but unlike in the expression of $|M_0^{(1)}] $ in \eq{OneLoop}, the lower limit of $\mb I^{(ij)}$ insertion is bounded below by the first (the most energetic) soft gluon transverse momentum, rendering it IR-finite.
In the next two lines, the lower limit of each of the $\mb I^{(ij)}$ insertions is set by the next soft gluon emission with index $n+m+1$ (not shown) and the transverse momentum evaluated in the dipole frame $\textcolor{Red}{(ij)}$ or $(\textcolor{Blue}{(n+m)}\textcolor{Red}{j})$ between which the virtual momentum is exchanged.
The upper limit is set by the latest soft gluon (with index $n+m$).

Interestingly, the diagram (c) specifies that the $k_T$ of the latest emission must be evaluated in the rest frame of its \textit{parent-dipole}, $\textcolor{Red}{(ij)}$. Because of this the individual dipole contributions $\mb d^{(0)}_{\textcolor{Red}{ij}} (\textcolor{Blue}{q_m})$ yield a distinct contribution and cannot be summed over to yield the tree level current $\mb J^{(0)}(\textcolor{Blue}{q_m})$ as in diagram (b), preventing the result from being written as a simple evolution equation.
This result can be contrasted with the work in \Ref{Neill:2018mmj}, where it was suggested that the correct ordering variable is \textit{always} the transverse momentum evaluated in the dipole frame of the collinear legs to which the virtual gluon is attached. While their statement is true for the diagrams of class (b) in \fig{AMFS}, it does not hold for diagrams of class (c) where the virtual gluon is exchanged between a soft real emission and a collinear leg. The analysis with two ordered soft gluon emissions further reveals additional exceptional cases like in \fig{AMFS}c.
We will see from the EFT analysis below that diagrams of class (c) arise as a result of an interesting \textit{memory effect}: as one successively moves between EFTs where the previous soft gluon is resolved and frozen to be a collinear direction, the new low energy EFT inherits the $k_T$ of the resolved emission in the parent dipole, $\textcolor{Blue}{q_{m\perp}^{(\textcolor{Red}{ij})}}$, as an additional, new \textit{hard scale} that restricts the transverse loop momentum whenever this resolved gluon is involved.

It is interesting to compare the AMFS result with resummation in SCET.
This story is straightforward for global measurements where the soft function is rendered finite via renormalization and it obeys an evolution equation. For non-global measurements, we can resort to the dressed gluon approximation~\cite{Larkoski:2015zka} where one performs an expansion in number of resolved emissions. This bears similarities with the AMFS result where the soft emissions are successively resolved, each soft emission sources the subsequent ones. In the dressed gluon approximation, an additional measurement must be performed at each stage to resolve yet softer emissions. Analogously, in the AMFS result, one keeps track of the transverse momentum of each additional soft gluon. In deriving \eq{AMFS} we will employ the same line of attack by matching through a sequence of EFT's where at each stage a soft emission is resolved.

\section{One soft gluon emission}
\label{sec:OneGluon}
We now discuss derivation of the AMFS result in \eq{AMFS} in the SCET framework. To pin-down the kinematic parameter that correctly constrains limits of the virtual loop momentum for a coherent cancellation of IR divergences we wish to evaluate the imaginary part of one-loop soft emission graphs, from which the real part can be obtained via analytical continuation using
\begin{align}
\im \pi \eps \tilde \delta_{ij} - \eps \ln \frac{\omega_{ij}}{\mu^2} = \eps \ln \frac{\mu^2}{-\omega_{ij}} \, .
\end{align}
We provide further justification in \sec{EFTnp1match} why logarithms involving $\omega_{ij}$ appear in the form shown on the right hand side.

The real part of one-loop virtual graphs can be expressed as a phase space integration, and it does not uniquely constrain the choice of the ordering variable. This is because any choice of variable $z^a \theta^b$ for $a > 0$ can serve as a UV or IR cutoff for the phase space integration. The imaginary $\im \pi$ terms, on the other hand, involve solely an integral over the transverse momentum, which is specified in a specific dipole frame. Hence, we will limit ourselves to calculating the imaginary part. Additionally, the imaginary parts are also easier to obtain than the real parts as they involve \textit{double cut} diagrams with both the longitudinal loop-momentum components constrained, as opposed to \textit{single cut} diagrams for the real part.

Our strategy will be to consider a sequence of effective field theories in order to describe amplitude for successively softer emissions. In each successive EFT we will have one extra soft resolved emission at lower virtuality than the previous one. By doing calculations with Glauber operators described below we will be able to obtain results for imaginary parts of the amplitudes quite efficiently.

\subsection{The EFT setup}
\label{sec:EFTsetup}
We consider the amplitude for $n$ hard partons which can be incoming/outgoing quarks, anti-quarks or gluons. The momentum of each of these $n$ hard partons defines a collinear direction $n_i$, and a momentum $k$ decomposed in these coordinates is given by
\begin{align}\label{eq:LCDef}
k^\mu = \big(k^+, k^-, k_{\perp}\big)_{n_i} \,, \qquad k^\mu = k^+ \frac{\bar n_i^\mu}{2} + k^- \frac{n_i^\mu}{2} + k_{\perp}^\mu \, ,
\end{align}
where $\bn_i$ is an auxiliary light like vector satisfying $\bn_i \cdot n_i = 2$. The collinear momenta $p_i^c$ and soft momenta $p_s$ satisfy the scaling
$$ p_i^c \sim Q (\lambda^2, 1, \lambda)_{n_i} \, , \qquad p_s \sim Q(\lambda,\lambda, \lambda) \, , $$
where $Q$ is the underlying hard scale and $\lambda \ll 1$ is the power counting parameter. We additionally demand that various $n$ directions are well resolved, such that $n_i \cdot n_j \sim 1 \gg \lambda$. In SCET, quark and gluon fields with these scalings are distinguished and represented as distinct soft and collinear fields. These fields interact each other via hard scattering operators or via Glauber potential operators.
The hard scattering in SCET is described via effective operators,
\begin{align}\label{eq:OnCompact}
O_n= \sum_{\Gamma} \int \Big(\prod_{i = 1}^n \df \omega_i\Big) \big[O_{n} \big(\{\omega_i,n_i\}\big) \big| {\cal C}_{n,\Gamma} \big( \{\omega_i \}\big)\big]\,
,
\end{align}
where $\omega_i \sim Q$ are the large momenta in direction $n_i$, and $\Gamma$ are possible Dirac structures. The $\omega_{ij}$'s introduced in \eq{OneLoop} are simply related to $\omega_i$ and $\omega_j$. We first choose the reference vectors $n$ and $\bn$ to lie along $p_i$ and $p_j$ momenta, such that
\begin{align}
n_i^\mu = \frac{p_i^\mu}{\kappa_{ij}} \,, \qquad n_j^\mu = \frac{p_j^\mu}{\kappa_{ij}} \, , \qquad \kappa_{ij}= \sqrt{\frac{p_i \cdot p_j}{2}} \, ,
\end{align}
where the factors $\kappa_{ij}$ ensure that $n_i \cdot n_j = 2$. Thus, we have
\begin{align}
p_i^\mu = \frac{n_i^\mu}{2} \omega_i \,, \qquad p_j^\mu = \frac{n_j^\mu}{2} \omega_j \, ,
\end{align}
which immediately gives
$$
\omega_{ij}= \omega_i \omega_j \, .
$$

The operators and the Wilson coefficients in \eq{OnCompact} are dual-vectors and vectors in the color space. We provide a detailed explanation of this notation in \app{Color}. The off-shell modes generated via interactions of soft and collinear fields can be integrated out via BPS field redefinition~\cite{Bauer:2001yt} of the collinear fields to obtain soft Wilson lines in directions $n_i$, such that
\begin{align}\label{eq:OnSCET}
O_n= \sum_{\Gamma} \int \Big(\prod_{i = 1}^n \df \omega_i\Big) \big[O^{(0)}_{n} \big(\{\omega_i,n_i\}\big) \big| \prod_{i=1}^n \mb S_{n_i} \,
\big| {\cal C}_{n,\Gamma} \big( \{\omega_i \}\big)\big]\,
.
\end{align}
The operator $O^{(0)}_{n}$ now consists solely of collinear fields. The expression of the Wilson lines for various cases are given in \eqs{WilsonLinesExplicit}{WilsonLinesExplicit2} and in \eq{WilsonLines}. Note that the superscript $(0)$ here is unrelated to any loop-expansion and simply denotes that the collinear fields in $O^{(0)}_{n}$ have no Lagrangian interactions (i.e. other than those through the hard scattering and Glauber operators) with the soft fields.

Next, the forward scattering process between these collinear legs and the soft emissions is mediated by the corresponding operators in the Glauber Lagrangian~\cite{Rothstein:2016bsq}:
\begin{align}\label{eq:OGSCET}
O_G(x) = e^{-\im x \cdot {\cal P}}\sum_{i,j} O_{n_i sn_j}^{ij} + e^{-\im x \cdot {\cal P}} \sum_{i} O^{ij}_{n_is} \subset {\cal L}_{G}^{{\rm II}(0)} (x) \, ,
\end{align}
where the superscripts $i,j$ specify the representation of the collinear legs. The $x$ dependence captures the long distance physics. The operator $O^{ij}_{n_i sn_j}$ describes forward scattering between two collinear partons flying in directions $n_i$ and $n_j$, with soft particles emitted at intermediate rapidity. The operator $O^i_{n_is}$ describes forward scattering process between an $n_i$-collinear and a soft parton. These operators thus contain soft Wilson lines as in \eq{OnSCET} and hence enter the calculation of the amplitude of soft gluon emissions. The Glauber operators are reviewed in \app{GlauberOps}.

The correspondence of QCD amplitudes with the combination of SCET amplitudes and Wilson coefficients for $n$ hard partons and arbitrary number of soft emissions is given by
\begin{align}\label{eq:QCD_EFT}
& \sum_{N = 0}^\infty (g \mu^\eps)^{{N}} \mb J ({q_N}) \ldots \mb J ({q_1})
\big| {\cal M} (p_1 , \ldots, p_n) \big ]
\\
&\qquad = \sum_\Gamma \int \Big(\prod_{i = 1}^n \df \omega_i\Big) \langle \{p_i\},\{q_j\} |\bigg({\rm T}\: O^{(0)}_{n} \big(\{\omega_i,n_i\}\big) \prod_{i=1}^n \mb S_{n_i} e^{\im \int \df^4 x' O_G(x') } \bigg)| 0 \rangle \big| {\cal C}_{n,\Gamma} \big( \{\omega_i \}\big)\big] \,.
\nn
\end{align}
We have expressed the matrix element as a matrix in the color space so as to more easily handle color mixing resulting from radiative corrections, as made explicit by the relation
\begin{align}
\sum_{\{a_i\}} \Big[ \{a_i\} \Big| O^{(0)}_{n} \big(\{\omega_i,n_i\}\big) = \big [O_n^{(0)}\big(\{\omega_i, n_i\}\big) \big |.
\end{align}

To obtain the imaginary parts of the one-loop graphs with soft emissions we will make use of the \textit{soft-Glauber correspondence} discussed in \Ref{Rothstein:2016bsq} which states that in the effective theory with Glauber modes, the graphs with Glauber attachments (to active partons) fully account for the $\im \pi$ terms once the zero bin has been subtracted from the soft graphs. Additionally, as noted in \Ref{Rothstein:2016bsq}, using of Glauber graphs provides an efficient way of obtaining the $\im \pi$ terms.

\subsection{Matching with no resolved soft emissions}
\label{sec:EFTn0R}
We will regulate the IR divergences via a gluon mass term, and the UV divergences in dimensional regularization.
The results derived in this section will be valid to ${\cal O}(\eps)$, and we will discuss higher order $\eps$ terms in \sec{EFTnp1match}. Lastly, the label Dirac structure $\Gamma$ on the Wilson coefficient $ \big| {\cal C}_{n,\Gamma} \big] $ in \eq{QCD_EFT} plays no important role in our calculations and hence we will suppress this dependence from hereon for simplicity.
Much of the results can be derived by restricting to the case of two outgoing hard partons, so that is where we will begin. At tree level, without any external soft emission, \eq{QCD_EFT} reads
\begin{align}
\left[ \{a_i\} \Big| {\cal C}_n^{(0)}(\{\omega_i\},\mu)\right] = \big[\{a_i\}\big| M_0^{(0)}\big] \, .
\end{align}
Consider first the one-loop graph without any soft real emission.
The imaginary part of full theory graph is given by
\begin{align}\label{eq:FullTheory1L0R}
\Im \left[\includegraphics[height=2.5cm,valign=c]{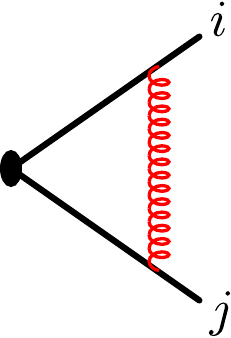}\right]_{\{a_i\}} =\quad \left [\includegraphics[height=2.5cm,valign=c]{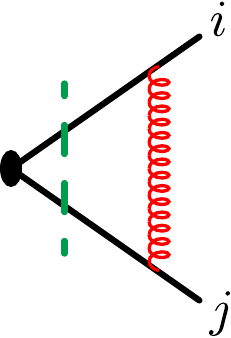}\right]_{\{a_i\}}
= \big[ \{a_i\}\big| \mb C^{(ij)}(m,\sqrt{\omega_{ij}}) \big|M_0^{(0)}\big] \, ,
\end{align}
and $|M_0]$ is the tree level hard matrix element.
The green line represents an \textit{Eikonal cut}~\cite{Catani:2008xa,Cutkosky:1960sp,Martinez:2016vur} that puts the partons $i$ and $j$ with momenta $p_i$ and $p_j$ onshell. Here we use a gluon mass $m$ as an infrared cutoff to regulate IR divergences, and have defined
\begin{align}\label{eq:CijDef}
\mb C^{(ij)}(a,b) \equiv
\frac{-\im \pi g^2}{8\pi^2} (\mb T_i \cdot \mb T_j) \int_{a^2}^{b^2} \frac{\df {\vec \ell_\perp}^{\,2}}{\vec \ell_\perp^{\,2}}
=(-\im \pi )
\frac{\alpha_s}{2\pi} (\mb T_i \cdot \mb T_j) \ln\Big(\frac{b^2}{a^2}\Big) \, ,
\end{align}
while the object $ \mb C^{(ij)}(a,b)$ itself involves a finite integral, and in case of \eq{FullTheory1L0R} finite after including the IR regulator, in SCET we will find UV divergent amplitudes where dimensional regularization will be the appropriate regulator. In this case, the $d$-dimensional UV-divergent version of \eq{CijDef} is given by:
\begin{align}\label{eq:CijmuDef}
\mb C^{(ij)} (m, \mu) \equiv (-\im \pi ) \frac{\alpha_s}{2\pi} (\mb T_i \cdot \mb T_j) \left(\frac{1}{\eps}+ \ln\frac{\mu^2}{m^2} \right) \, ,
\end{align}
where the gluon mass $m$ regulates the IR divergence in this otherwise scaleless diagram.
Thus, we will understand $\mb C^{(ij)}$ to mean \eq{CijmuDef} when $\mu$ appears as one of the argument, otherwise for physical cutoffs \eq{CijDef} will be employed. We note that, when using dimensional regularization, the imaginary parts we obtain below using cut diagrams or Glauber exchanges will only be accurate to ${\cal O}(\eps)$. We address this issue in detail below in \sec{EFTnp1match}.

The same graph can also be calculated in SCET where the two hard partons are replaced by two collinear partons. The imaginary part in the SCET graph is obtained by considering a Glauber exchange between the two legs.
The final result will be given by combination of the Wilson coefficient and the SCET matrix element:
\begin{align}
\Im \left[\includegraphics[height=2.5cm,valign=c]{QCD_1L0R_nocut.pdf}\right]_{\{a_i\}} &=
\Im \left[
[ \{a_i\}|
\left(\includegraphics[height=1.5cm,valign=c]{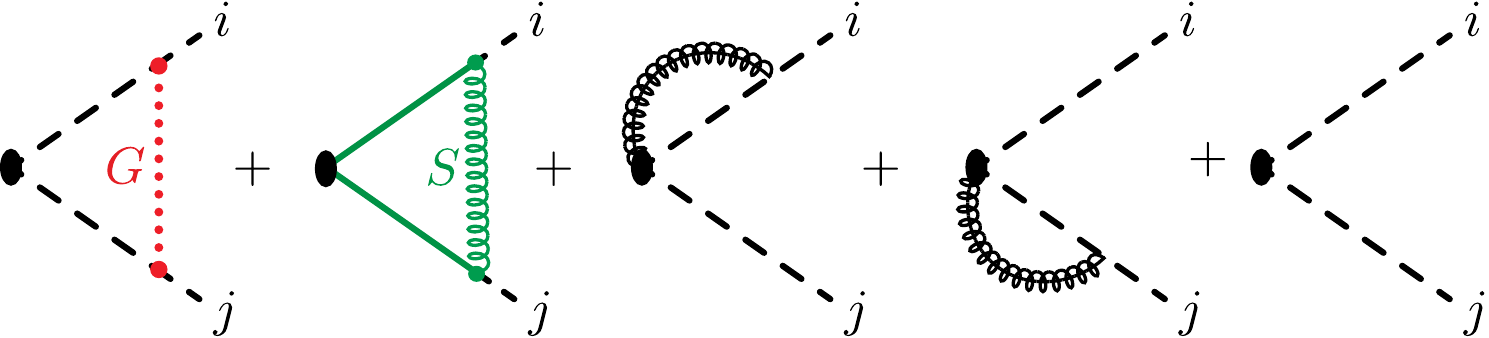} \times Z_\xi\right)
\big| {\cal C}_n (\{\omega_i\},\mu) \big]
\right] \nn \\
& = \sum_{\{b_i\}} \: [ \{a_i\}| \left(\includegraphics[height=1.5cm,valign=c]{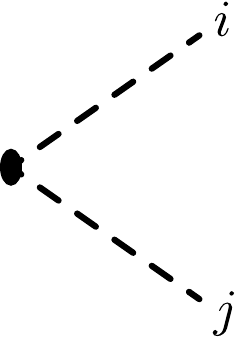}\right) | \{b_i\} ] \times \Im \left[\{b_i\}\Big | {\cal C}^{(1)}_n (\{\omega_i\},\mu) \right]
\nn\\
&
\quad+ \sum_{\{b_i\}}\:
[ \{a_i\}| \left(\includegraphics[height=1.5cm,valign=c]{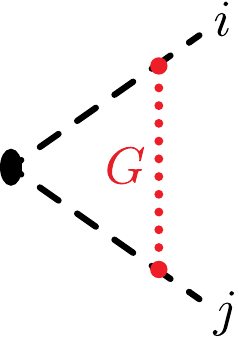}\right) | \{b_i\} ] \times \Re\left[\{b_i\}\Big | {\cal C}_n^{(0)}(\{\omega_i\},\mu)\right] \, .
\end{align}
At one-loop accuracy, the SCET matrix element with the Glauber exchange accounts for the imaginary part and is given by the operator
\begin{align}\label{eq:G0ij}
G_0 \equiv
\sum_{i=1}^n\sum_{j=1}^{i-1} \left(
\includegraphics[height=2.5cm,valign=c]{SCET_1L0R.pdf}\right) = \sum_{i=1}^n\sum_{j = 1}^{i -1 } \mb C^{(ij)}(m,\mu) \, ,
\end{align}
with the details of the calculation provided in \app{1L0R}.
Hence, the imaginary part of the one-loop Wilson coefficient is given by
\begin{align}\label{eq:ImCn}
\Im \left |{\cal C}^{(1)}_n (\{\omega_i\},\mu) \right] = \sum_{i=1}^n\sum_{j = 1}^{i-1} \mb C^{(ij)} (\mu, \sqrt{\omega_{ij}}) \big | M_0^{(0)}\big]\, .
\end{align}

\subsection{Amplitude for single soft gluon emission}
\begin{figure}[t]
\centering
\includegraphics[width=0.5\textwidth]{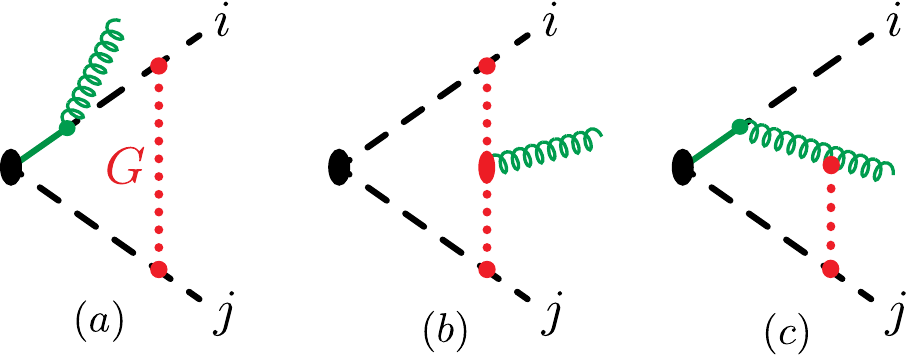}
\caption{Diagrams that give imaginary part of one-loop one real emission amplitude. Additional diagrams are obtained by flipping $i\lra j$ in (a) and (c).}
\label{fig:1L1R}
\end{figure}

Having fixed the Wilson coefficient we can now evaluate amplitude for single soft emission. The result will again be given by combination of the imaginary part of the one-loop Wilson coefficient and one-loop graphs with Glauber exchanges shown in \fig{1L1R}. The results for these graphs are given by
\begin{align}
\includegraphics[height=2.5cm,valign=c]{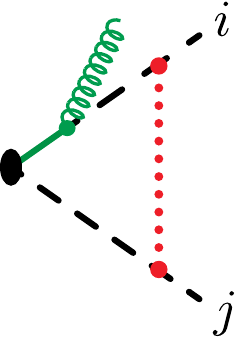}
= g\, \mb C^{(ij)} (m, \mu)\bigg( \mb T_i^{a_1} \frac{\veps^*(q_1) \cdot n_i}{{q_1} \cdot n_i} \bigg) \, ,
\end{align}
such that summing over all the attachments of the soft gluon and all the pairwise Glauber exchanges yields
\begin{align}\label{eq:G1a}
G_{1(a)} = g \, \sum_{i=1}^n\sum_{j = 1}^{i-1} \mb C^{(ij)}(m, \mu) \: \mb J^{(0)} ({q_1}) \, .
\end{align}

Next, we have the Lipatov vertex graph in \fig{1L1R}b. As worked out in \app{Lipatov}, the graph involves the following transverse momentum integrals,
\begin{align}\label{eq:LipatovSwitch}
\frac{\big(q_{1\perp}^{(ij)}\big)^2}{2} \int \frac{\mu^{2\eps} \: \dfbar^{d-2} \ell_\perp}{(\ell_\perp^2 -m^2) \,[ (\ell_\perp+ q_{1\perp}^{(ij)})^2 -m^2]} - \int \frac{\mu^{2\eps} \: \dfbar^{d-2} \ell_\perp}{(\ell_\perp^2 -m^2)} \, ,
\end{align}
where the transverse momentum ${q_{\perp}^{(ij)}}$ is defined in the $(ij)$ dipole frame defined above in \eq{qperpijDef}.
This combination, as we now show, results in an IR cutoff of $q_{1\perp}^{(ij)}$ on the loop momentum $\ell_\perp$.
Writing $q_{1\perp}^{(ij)}$ as $q_{1\perp}$ and adding $k_\perp \cdot q_{1\perp}$ in the numerator we find
\begin{align}\label{eq:perpInteg}
\frac{q_{1\perp}^2}{2} \int \frac{\mu^{2\eps} \: \dfbar^{d-2} \ell_\perp}{(\ell_\perp^2 -m^2) \,[ (\ell_\perp+ q_{1\perp})^2 -m^2]}
&= \int \frac{\mu^{2\eps} \:\dfbar^{d-2} \ell_\perp}{(\ell_\perp^2 -m^2)}\frac{q_{1\perp}\cdot (\ell_\perp+ q_{1\perp})}{[(\ell_\perp+ q_{1\perp})^2 - m^2]} \\
&= \int \frac{\mu^{2\eps} \:\dfbar^{d-2} \ell_\perp}{(\ell_\perp^2 -m^2)} \frac{1 + \frac{|\vec \ell_\perp|}{|\vec q_{1\perp}| }\cos \phi}{1 + \frac{|\vec \ell_\perp|^2 +m^2}{|\vec q_{1\perp}| } + 2\frac{|\vec \ell_\perp|}{|\vec q_{1\perp}| }\cos \phi} \, .\nn
\end{align}
Using
\begin{align}
\int_0^{2\pi} \frac{\df \phi}{2\pi} \frac{1}{a+ b\cos \phi} = \frac{1}{\sqrt{a^2 - b^2}} \, ,
\qquad
\int_0^{2\pi} \frac{\df \phi}{2\pi} \frac{b\cos \phi}{a+ b\cos \phi} = 1 - \frac{a}{\sqrt{a^2 - b^2}} \, ,
\qquad (a>b)\,,
\end{align}
we have
\begin{align}\label{eq:ThetaFunc}
\int_0^{2\pi} \frac{\df \phi}{2\pi} \frac{1 + \frac{|\vec \ell_\perp|}{|\vec q_{1\perp}| }\cos \phi}{1 + \frac{|\vec \ell_\perp|^2 +m^2}{|\vec q_{1\perp}| } + 2\frac{|\vec \ell_\perp|}{|\vec q_{1\perp}| }\cos \phi} = \Theta ( | \vec q_{1\perp} | - | \vec \ell_\perp| ) + {\cal O}(m^2)\, .
\end{align}
This implies that the integral in \eq{perpInteg} is UV finite, and hence can be carried out in 4 dimensions. Subtracting the second term in \eq{LipatovSwitch} turns $q_{1 \perp}$ into lower bound and after some algebra we find
\begin{align}\label{eq:G1b0}
G_{1(b)} \equiv
\sum_{i=1}^n\sum_{j=1}^{i-1} \left(\includegraphics[height=2.5cm,valign=c]{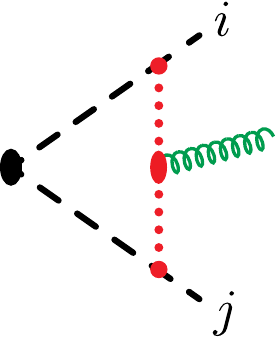} \right)
&= g\, \sum_{i=1}^n\sum_{j=1}^{i-1}\Big[ \mb d^{(0)}_{ji} (q_1) , \mb C^{(ij)} ( q_{1\perp}^{(ij)}, \mu )\Big] \, .
\end{align}
We can make $i\lra j$ symmetry in this result explicit by noting that the commutator $[\mb d^{(0)}_{jk}(q_1), \mb C^{(ij)}]$ vanishes for $k \neq i$. Including these terms we obtain
\begin{align}\label{eq:G1b}
G_{1(b)}
&=g\, \sum_{i=1}^n\sum_{j=1}^{i-1} \Big[ \mb J^{(0)} (q_1) ,\: \mb C^{(ij)} ( q_{1\perp}^{(ij)}, \mu )\Big] \, .
\end{align}
The Lipatov vertex graph alone thus implements the ``switch mechanism'' identified in \Ref{Angeles-Martinez:2015rna}. Physically, this means that the loop transverse momentum in graphs in \eq{G1b0} must be at least $q_{1\perp}^{(ij)}$ so as to account for production of the soft gluon.
Adding this result to that of $G_{1(a)}$ in \eq{G1a} we find
\begin{align}
G_{1(a)} + G_{1(b)} = g\, \sum_{i=1}^n\sum_{j=1}^{i-1} \Big(
\mb C^{(ij)} (m, q_{1\perp}^{(ij)} ) \mb J^{(0)} (q_1)
+ \mb J^{(0)} (q_1) \mb C^{(ij)} ( q_{1\perp}^{(ij)}, \mu )
\Big) \, .
\end{align}

Lastly, the rescattering graph is given by (see \app{Rescatter})
\begin{align}\label{eq:G1c}
G_{1(c)} &\equiv \sum_{i = 1}^n \sum_{j=1}^{i-1}\left( \includegraphics[height=2.5cm,valign=c]{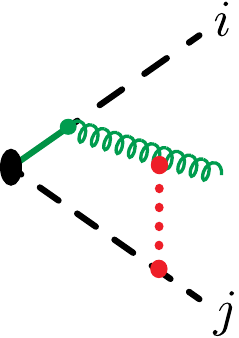}
+ (i\lra j)\right)
=g\, \sum_{i = 1}^n \sum_{\mathclap{\substack{j = 1\\ j\neq i}}}^n
\mb C^{(q_1 j )}(m , q_{1\perp}^{(ij)})\, \mb d^{(0)}_{ji} (q_1) \,.
\end{align}
Interestingly, we see that the diagram does not contribute to the UV pole. It involves only the integral in \eq{perpInteg} and is thus bounded above by $q_{1\perp}^{(ij)}$. Additionally, as noted above, since the loop momentum depends on the transverse momentum in the dipole frame of the parent, one cannot sum over $i$ to recover the tree level current.

Finally, the imaginary part of the one-loop one-real emission amplitude is given by sum of these three contributions, as well as the one-loop Wilson coefficient in \eq{ImCn}:
\begin{align}\label{eq:ImMn}
&\Im \bigg (\Big[ \{a_i\}, C_1\Big| \langle( q_1, \veps_1), \{p_i\} | O_n\big(\{\omega_i, n_i\}\big) | 0 \rangle \Big | {\cal C}_n \big(\{\omega_i \} , \mu\big)\Big ]\bigg)^{(1)}
\\
&
=\Big[ \{a_i\}, C_1 \Big| \bigg( g\, \mb J^{(0)} (q_1) \times \Im \left |{\cal C}^{(1)}_n (\{\omega_i\},\mu) \right]
+G_{1(a+b+c)}(m,q_1, \mu)\times \Re\left|{\cal C}_n^{(0)}(\{\omega_i\},\mu)\right]
\bigg)
\nn \\
& =g \Big[ \{a_i\}, C_1 \Big| \Bigg(
\mb J^{(0)} (q_1) \sum_{i = 1}^n \sum_{j=1}^{i-1}\mb C^{(ij)} (q_{1\perp}^{(ij)},\sqrt{ \omega_{ij}}) \nn \\
&\qquad \qquad
+ \sum_{i = 1}^n \bigg[ \sum_{j=1}^{i-1} \mb C^{(ij)}(m, q_{1\perp}^{(ij)}) \, \mb J^{(0)} (q_1)
+ \sum_{\mathclap{\substack{j = 1\\ j\neq i}}}^n
\mb C^{(q_1 j )}(m , q_{1\perp}^{(ij)}) \, \mb d^{(0)}_{ji} (q_1) \bigg]
\Bigg) \big|M_0^{(0)}\big] \, .\nn
\end{align}
Including the Wilson coefficient in \eq{ImCn} appropriately turns the $\mu$ scale to the high scale $Q$.
This result agrees with the calculation in full theory presented in \Ref{Angeles-Martinez:2015rna}.

\section{Two soft gluon emissions}
\label{sec:twoGluon}

We now consider the case where there is an additional soft gluon $q_2$, yet softer than $q_1$, such that $q_2 \sim \rho Q$ and $q_1 \sim \lambda Q$, where $\rho$ is a power counting parameter parametrically smaller than $\lambda$, $\rho \ll \lambda \ll 1$, and $Q$ is a generic hard scale. Our strategy to calculate matrix elements with this additional gluon will be to match the EFT with the $q_1$ soft gluon to an EFT at yet lower energies where this soft gluon is considered resolved, and the virtualities of $q_1$ and the other hard partons are only allowed to fluctuate to scales $\mu^2 \sim Q^2 \rho^2$, instead of $\mu^2 \sim Q^2 \lambda^2$ as in the original EFT. By doing so, we will find that the calculation of one-loop, double soft emission amplitude in the ordered limit significantly simplifies.

\subsection{Operators in the low energy EFT}
\label{sec:EFTnp1Def}

For sake of brevity, let us refer to the original effective theory with $n$ hard partons as EFT$_n$, and the low energy EFT described above as \eftnp
The appropriate current in EFT$_{n+1}$ can be expressed as
\begin{align}
O_{n+1}^{\rm hard \,scatter}
= \int \Big(\prod_{i = 1}^{n+1} \df \omega_i \Big)
\big[O_{n+1}( \{ \omega_1 , n_1, \omega_i ,n_i\}) \big | {\cal C}_{n+1} ( \{\omega_1, \omega_i \},\mu) \big] \, .
\end{align}
The operator $O_{n+1}$ includes gluon $q_1$ as an additional collinear direction, that can source additional soft modes in EFT$_{n+1}$. Here $n_1^\mu$ is a light-like vector in the direction of $q_1^\mu$, such that when taken as a reference vector, the momentum $q_1^\mu$ scales as
\begin{align}\label{eq:q1Comp}
(q_1 \cdot n_1, q_1 \cdot \bar n_1, q_{1\perp} ) \sim \omega_1 (\rho^2, 1, \rho) \, .
\end{align}
We can use any auxiliary vector $\bar n_1$ satisfying $n_1 \cdot \bar n_1 \sim {\cal O}(1)$ to decompose the momentum.
Here $\omega_1$ is the \textit{hard scale} associated with the resolved soft gluon $q_1$. We will identify this scale below.
Finally, we will use the notation $\omega_{n+1} = \omega_1$, $n^\mu_{n+1} = n^\mu_1$ etc.

We can write down the appropriate $O_{n+1}$ by demanding that the additional particle have quantum numbers of gluon, and the operator be Lorentz invariant, gauge invariant under collinear and soft gauge transformations~\cite{Bauer:2001ct}, as well as reparametrization invariant~\cite{Manohar_2002}. The appropriate gauge invariant building block for collinear gluons is the field ${\cal B}_{n_1\perp}^{\mu a_1}$~\cite{Bauer:2001ct}, and the representation in which this field is expressed depends on which of the $n$ hard partons sourced it. Thus, the operator satisfying these constraints is given by
\begin{align}\label{eq:Onp1}
\big[O_{n+1}( \{\omega_1, n_1, \omega_i ,n_i\}) \big | \equiv
\big [O_n\big(\{\omega_i, n_i\}\big) \big | \Bigg[g\sum_{i = 1}^n \frac{n_i \cdot {\cal B}_{n_1\perp, \omega_1}^{a}}{n_i \cdot q_1} \mb T_i^a\Bigg] \, .
\end{align}
Here, the `$\perp$' symbol implies that the vector is perpendicular to $n_1$. It is convenient to take $\bar n_1$ in \eq{q1Comp} to be along one of the hard partons in the direction $n_j$, such that any momentum $k^\mu$ decomposed in this frame is given by
\begin{align}\label{eq:kmu1j}
k^\mu \big|_{(1j)}= \frac{n_j^\mu}{2} k^{+(1j)} + \frac{n_1^\mu}{2} k^{-(1j)} + k_{\perp(1j)} \, ,
\quad n_1^\mu = \frac{q_1^\mu}{\sqrt{\kappa_{1j}}} \, , \quad n_j^\mu = \frac{p_j^\mu}{\sqrt{\kappa_{1j}}} \,,
\quad \kappa_{1j} \equiv \frac{q_1 \cdot p_j}{2} \, .
\end{align}
At lowest order, we have
\begin{align}\label{eq:BnperpExp}
{\cal B}_{n_1\perp}^{\mu(1j)a} &= A_{n_1\perp(1j),k}^\mu - \frac{k_{\perp(1j)}^\mu}{n_j \cdot k} n_j \cdot A_{n_1,k} + \ldots \nn \\
&= g_{\perp(1j)}^{\mu\nu} \left((A_{n_1,k} )_\nu- k_\nu \frac{n_j\cdot A_{n_1,k}}{n_j\cdot k} \right) + \ldots \, .
\end{align}
This relation is derived below in \app{FeynGlauber}.

Next, we note that while the representation of $\mb T_i$ in each term depends on the $i^{\rm th}$ sector, upon BPS field redefinition, \eq{Onp1} will always result in an additional Wilson line $\mb S_{n+1} = \mb S_{q_1}$ in the adjoint representation for each term, such that
\begin{align}
\big[O_{n+1}( \{\omega_1, n_1, \omega_i ,n_i\}) \big | =
\big [O_n^{(0)}\big(\{\omega_i, n_i\}\big) \big | \Bigg[g\sum_{i = 1}^n \frac{n_i \cdot {\cal B}_{n_1\perp, \omega_1}^{(0)a}}{n_i \cdot q_1} \mb T_i^a\Bigg]\Big( \prod_{i = 1}^n \mb S_{n_i} \Big) \mb S_{q_1} \, ,
\end{align}
where the color matrix $\mb T_{q_1}$ in $\mb S_{q_1}$ is in the adjoint representation.

Likewise, in \eftnp, we will update the Glauber operators in \eq{OGSCET} to include additional Glauber interactions involving the $n+1$ gluon as a collinear mode. The fact that the additional resolved gluon is no longer treated as a soft mode in \eftnp will have important consequences for the one-loop calculations in this theory. For example, it can no longer be sourced by the Lipatov vertex between two other collinear sectors, as in the diagram \fig{1L1R}b, and Glauber exchanges between this gluon and other softer gluons will now be allowed. Additionally, in \eftnp soft emissions (yet softer than $q_1$) can be produced via the Wilson lines in mid rapidity operators $\{O_{n_{1}sn_j}\}$ involving $q_1$ and any other collinear sector. These cases are shown in \figs{n1G0}{1L2R}.

\subsection{Matching to the low energy EFT}
\label{sec:EFTnp1match}

As a first step, let us check that the operator in \eq{Onp1} correctly reproduces the tree level current for $q_1$ emission:
\begin{align}
&\Big[\{a_i\}, C_1 \Big| \langle( q_1, \veps_1) ,\{p_i\} | O_{n+1}( \{ \omega_1, n_1, \omega_i ,n_i\}) | 0 \rangle \Big | {\cal C}_{n+1} (\{ \omega_1, \omega_i \},\mu) \Big]^{(0)} \\
&\qquad =g \Big[ \{a_i\}, C_1 \Big| \sum_{i = 1}^n \frac{n^\mu_{i\perp(1j)}}{n_i\cdot q_1} \left(\veps^*_\mu(q_1) - q_{1\mu} \frac{n_j\cdot \veps^*(q_1)}{n_j\cdot q_1}\right) \mb T_i \Big | {\cal C}_{n+1} \Big]^{(0)} \nn \\
&\qquad = g \Big[ \{a_i\}, C_1 \Big| \sum_{i = 1}^n \Big[\left(\frac{n_i\cdot \veps^*(q_1)}{n_i\cdot q_1} - \frac{n_j\cdot \veps^*(q_1)}{n_j\cdot q_1}\right)
\nn \\
&\qquad \qquad \qquad - \frac{n_i \cdot n_j }{n_i\cdot q_1} \left(n_1 \cdot \veps^*(q_1) -\frac{ n_1 \cdot q_1 \,n_j\cdot \veps^*(q_1)}{n_j\cdot q_1}\right)\Big] \mb T_i \Big | {\cal C}_{n+1} \Big]^{(0)} \nn \\
&\qquad =g \Big[ \{a_i\}, C_1 \Big| \sum_{i = 1}^n \mb d^{(0)}_{ji}(q_1) \Big | {\cal C}_{n+1} \Big]^{(0)} (1 + {\cal O}(\rho^2)) =
g\Big[ \{a_i\}, C_1 \Big| \mb J^{(0)} (q_1) \Big | {\cal C}_{n+1} \Big]^{(0)}
\nn \, ,
\end{align}
and hence $|{\cal C}_{n+1}^{(0)}] = |M_0]$ yields natural normalization.

\begin{figure}[t]
\centering
\includegraphics[width=0.5\textwidth]{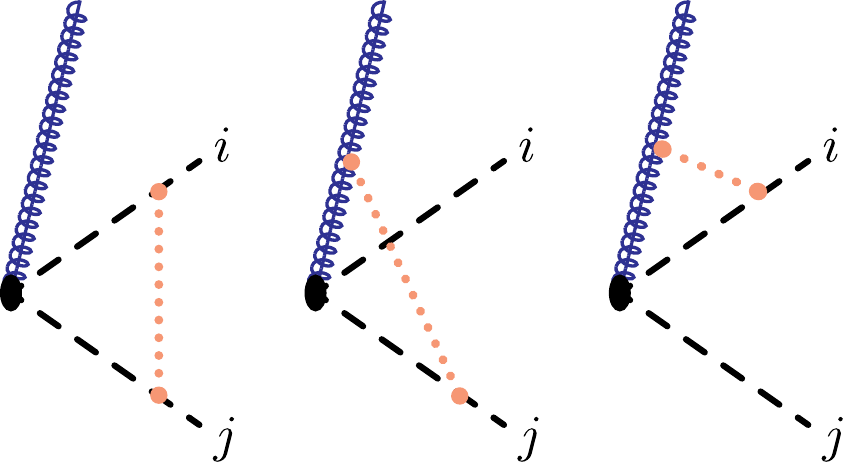}
\caption{One-loop Glauber exchange diagrams in low energy EFT without any additional soft emission. The $q_1$ gluon, while soft in the full theory, has now become a collinear mode. The Glauber exchanges shown here correspond to lower transverse momentum kicks.}
\label{fig:n1G0}
\end{figure}

Next, we will determine (the imaginary part of) the Wilson coefficient ${\cal C}_{n+1}$ at one-loop in an ordered form by calculating the one-loop onshell matrix element for $q_1$ emission in \eftnp using \eq{Onp1} and equating it with the expression obtained earlier in \eq{ImMn}. This will also allow us to identify the hard scale associated with the soft emission $q_1$. In \eftnp, the imaginary part of the matrix element is simply given by summing over all the pairwise Glauber exchanges shown in \fig{n1G0},
\begin{align}\label{eq:q1AmpLow}
&\Im \Big[\{a_i\}, C_1 \Big| \langle( q_1, \veps_1) ,\{p_i\} | O_{n+1}( \{ \omega_1, n_1, \omega_i ,n_i\}) | 0 \rangle \Big | {\cal C}_{n+1} (\{ \omega_1, \omega_i \},\mu) \Big]^{(1)} \\
&\qquad = g\,
\Big[\{a_i\}, C_1 \Big| \bigg(\sum_{i=1}^n\sum_{j=1}^{i-1} \mb C^{(ij)}(m, \mu) + \sum_{i = 1}^n \mb C^{(q_1 i)}(m, \mu) \bigg) \mb J^{(0)}(q_1) \big|M_0^{(0)}\big] \nn \\
&\qquad + g \Big[\{a_i\}, C_1 \Big| \mb J^{(0)} (q_1) \times \Im \Big| {\cal C}^{(1)}_{n+1} \Big]
\nn
\, .
\end{align}
Comparing with \eq{ImMn} we see that dependence on the infrared cutoff $m$ cancels on both the sides, consistent with expectation that $|{\cal C}_{n+1}]$ must only encode UV physics. Thus, we find
\begin{align}\label{eq:Cnp1}
\mb J^{(0)} (q_1) & \Im \left|{\cal C}^{(1)}_{n+1} ( \{q_{1\perp}^{(ij)},\omega_i\},\mu) \right] = \mb J^{(0)}(q_1) \sum_{i=1}^n\sum_{j=1}^{i-1} \mb C^{(ij)}(q_{1\perp}^{(ij)},\sqrt{\omega_{ij}}) \big|M_0^{(0)}\big] \\
&\qquad + \sum_{i = 1}^n \bigg( \sum_{j=1}^{i-1} \mb C^{(ij)}(\mu, q_{1\perp}^{(ij)}) \mb J^{(0)} (q_1)
+ \sum_{\mathclap{\substack{j = 1\\ j\neq i}}}^n \mb C^{(q_1 i)}( \mu, q_{1\perp}^{(ij)}) \mb d^{(0)}_{ij} (q_1)
\bigg) \big|M_0^{(0)}\big] \nn \, .
\end{align}
In the first line the previous hard matching corrections from \eftn to QCD is included, whereas the second line it is the $\omega_1 = q_{1\perp}^{(ij)}$ that sets the hard scale for \eftnp. Interestingly, using \eq{J0} this relation can also be expressed as
\begin{align}\label{eq:dijUV}
&\mb J^{(0)}(q_1) \Im\bigg( \left| {\cal C}^{(1)}_{n+1} (\{q_{1\perp}^{(ij)},\omega_i\}, \mu) \right]-\left| {\cal C}^{(1)}_n (\{\omega_i\},\mu) \right] \bigg)\\
&= \sum_{i=1}^n\sum_{j=1}^{i-1}\left( -\Big[\mb J^{(0)}(q_1), \mb C^{(ij)}(\mu, q_{1\perp}^{(ij)}) \Big]
+ \mb C^{(q_1 i)} (\mu, q_{1\perp}^{(ij)}) \mb d^{(0)}_{ij}(q_1) + \mb C^{(q_1 j)} (\mu, q_{1\perp}^{(ij)}) \mb d^{(0)}_{ji}(q_1) \right)\big|M_0^{(0)}\big] \nn
\, ,
\end{align}
such that the left hand side now corresponds to the Wilson coefficient for matching between the two EFTs:
\begin{align}
\big| \Delta {\cal C}_{n+1} (\{q_{\perp}^{(ij)}\}, \mu) \big] \equiv\big| {\cal C}_{n+1} (\{q_{1\perp}^{(ij)},\omega_i\}, \mu) \big]- \big | {\cal C}_n (\{\omega_i\},\mu) \big]\, .
\end{align}
Since the matrix elements in \eftn do not depend on $\omega_i$'s, the matching then only involves the transverse momenta of $q_1$ in various dipole frames. Next, we note that if we use dimensional regularization to regulate IR divergences, then the loop contributions to the onshell matrix element $\langle( q_1, \veps_1), \{p_i\} | [O_{n+1}( \mu) \big | 0\rangle$ in \eq{q1AmpLow} being scaleless integrals simply vanish in \eftnp~\cite{Becher:2009kw}, such that the condition
\begin{align}
\mb J^{(0)}(q_1) \left|\Delta {\cal C}^{\rm bare}_{n+1} (\{q_{1\perp}^{(ij)}\}, \mu ) \right]
= \langle( q_1, \veps_1), \{p_i\} | O_{n}( \mu) \big | 0\rangle \big|M_0^{(0)}\big] = \mb J (q_1) \big|M_0^{(0)}\big]\Big|_{\eps_{\rm IR} \ra \eps_{\rm UV}}
\, .
\end{align}
is satisfied to all orders. Here the right hand side is the loop expanded soft gluon emission operator with the IR divergences $\eps_{\rm IR}$ now interpreted as $\eps_{\rm UV}$. Specifically, at one loop we have
\begin{align}\label{eq:DeltaC}
\mb J^{ (0)}(q_1) \left| \Delta {\cal C}^{\rm bare(1)}_{n+1} (\{q_{1\perp}^{(ij)}\}, \mu ) \right] = \mb J^{(1)} (q_1) \big |M_0\big]= \sum_{i = 1}^n \sum_{j=1}^{i-1} \mb d^{(1)}_{ij} (q_1)\big|M_0^{(0)}\big]\, .
\end{align}
This relation in conjunction with \eq{dijUV} is in direct correspondence with the observation made in \Ref{Angeles-Martinez:2015rna} that the one-loop emission operator $\mb d_{ij}^{(1)}(q_1)$ satisfies the following identity:
\begin{align}\label{eq:dijIR}
\mb d_{ij}^{(1)}(q_1) \approx -\Big[ \mb J^{(0)}(q_1), \mb I^{(ij)} (0, q_{1\perp}^{(ij)}) \Big] + \mb I^{(q_1 i)} (0, q_{1\perp}^{(ij)}) \mb d^{(0)}_{ij}(q_1) + \mb I^{(q_1 j)} (0, q_{1\perp}^{(ij)}) \mb d^{(0)}_{ji}(q_1) \, .
\end{align}
Note that, in contrast to \eq{dijUV}, here both the sides are IR divergent as these correspond to amplitudes in the renormalized full theory. The approximate equality in \eq{dijIR} denotes that the relation is valid to ${\cal O}(\eps)$ when the IR divergences are regulated in dimensional regularization.
It was, however, noted in \Ref{Angeles-Martinez:2015rna} that the relationship in \eq{dijIR} becomes exact when $1 + \im \pi \eps$ is replaced by $\cos(\pi \eps) + \im \sin (\pi \eps)$. We can now understand this by relating it to \eqs{dijUV}{DeltaC}. Because $|\Delta {\cal C}_{n+1}]$ is a Wilson coefficient between \eftnp and \eftn, it must depend on the combination $ (-(q_{1\perp}^{(ij)})^2 - \im 0)$ such that the relation
\begin{align}
\Im \Big[(-(q_{1\perp}^{(ij)} )^2- \im 0)^{-1-\eps}\Big] = - \sin (\pi\eps) \big[(q_{1\perp}^{(ij)} )^2\big]^{-1-\eps} \, .
\end{align}
accounts for the higher order $\eps$ terms that are not captured through Glauber exchanges or cut diagrams. This also justifies our reasoning for obtaining the real part via analytical continuation.

\subsection{Adding another soft emission}
\label{sec:AddSofter}
\begin{figure}[t]
\centering
\includegraphics[width=0.9\textwidth]{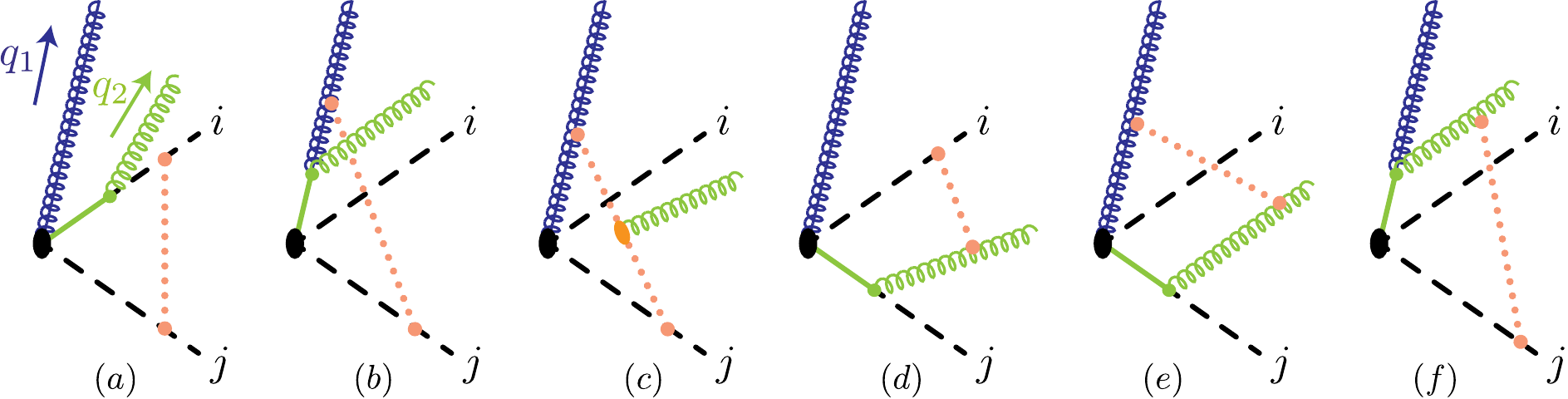}
\caption{One-loop Glauber exchange diagrams in low energy EFT with an additional soft emission. The solid line represents an offshell mode that has been integrated out.}
\label{fig:1L2R}
\end{figure}

Having determined the Wilson coefficient for the low energy theory, we are now equipped to evaluate diagrams with one additional soft emission. In this theory, the $q_1$ gluon behaves completely analogously to other hard partons, and hence the analysis of the previous case can be recycled. The tree level graphs are given by
\begin{align}\label{eq:tree2R}
\sum_{i = 1}^n\left( \includegraphics[height=2.8cm,valign=c]{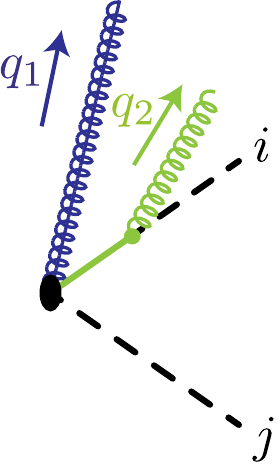} \right) \quad + \quad
\includegraphics[height=2.8cm,valign=c]{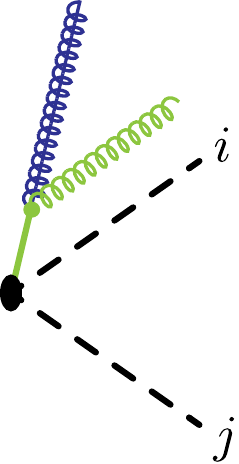}
= g^2 {\mb J}^{(0)}_2(q_2, q_1) \: \tilde{\mb J}^{(0)} (q_1) \, ,
\end{align}
where
\begin{align}\label{eq:J20Def}
\mb J^{\mu(0)}_2(q_2,q_1) = \tilde{\mb J}^{(0)\mu}(q_2) + \mb T_{q_1} \frac{q_1^\mu}{q_1\cdot q_2} \, , \qquad
\tilde{\mb J}^{(0)\mu}(q) \equiv \sum_{k = 1}^n \mb d_{jk}^{(0)\mu} (q) \, ,
\end{align}
and we use the shorthand
\begin{align}
{\mb J}^{(0)}_2(q_2, q_1) \equiv \veps^*_\mu(q_2) \mb J^{\mu(0)}_2(q_2,q_1) \, \qquad \tilde{\mb J}^{(0)} (q_1) \equiv \veps^*_\mu(q_1) \tilde{\mb J}^{(0)\mu}(q) \, .
\end{align}
We have used the symbol $\tilde{\mb J}^{(0)\mu}$ to distinguish it from the tree level $n$-parton current $\mb J^{(0)\mu}$ in \eq{J0} since the color conservation now applies to $(n+1)$-parton soft current, $\mb J_2^{(0)}(q_2, q_1)$.
In \fig{1L2R} we show some of the diagrams that enter the calculation of the imaginary part of the matrix element. Note that, unlike the more energetic gluon $q_1$, the softer emission $q_2$ can be produced via the Lipatov vertex. The rescattering graph in \fig{1L2R} involving exchange between the two gluon emissions alone completely accounts for the previously-termed case of ``soft gluon cuts'' in \Ref{Angeles-Martinez:2015rna}.

Graphs where the Glauber exchange is decoupled from the soft gluon emission (\fig{1L2R}a,b) sum up to the following expression:
\begin{align}\label{eq:G2a}
G_{2(a,b)} &=g^2 \, \sum_{i=1}^n\bigg[ \mb C^{(q_1i)}(m, \mu) + \sum_{j=1}^{i-1} \mb C^{(ij)}(m, \mu) \bigg] \mb J^{(0)}_2(q_2, q_1) \tilde {\mb J}^{(0)} (q_1) \nn \\
&= g^2\, \sum_{i=1}^{n+1} \sum_{j=1}^{i-1} \mb C^{(ij)}(m, \mu) \mb J^{(0)}_2(q_2, q_1) \tilde {\mb J}^{(0)} (q_1)
\, .
\end{align}
Next, the Lipatov vertex graphs are given by a direct generalization of \eq{G1b0}, such that
\begin{align}
G_{2(c)} = g^2 &\sum_{i = 1}^{n}
\Bigg( \Big[ \mb d^{(0)}_{iq_1} (q_2) , \mb C^{(q_1 i)} ( q_{2\perp}^{(q_1 i)}, \mu )\Big] + \sum_{j=1}^{i-1} \Big[ \mb d^{(0)}_{ji} (q_2), \mb C^{(ij)} ( q_{2\perp}^{(ij)}, \mu )\Big]\Bigg)
\tilde {\mb J}^{(0)} (q_1) \, .
\end{align}
The first term corresponds to the new contribution of the Lipatov vertex between $q_1$ and $j$, shown in \fig{1L2R}c.
As before, we can add additional vanishing commutators $[ \mb d^{(0)}_{ik}(q_2), \mb C^{(q_1i)}(q_{2\perp}^{(q_1i)}, \mu)]$ for $k\neq n+1$ in the first line and for $k \neq i$ in the second, such that
\begin{align}
G_{2(c)} = g^2\, &\sum_{i = 1}^{n+1} \sum_{j=1}^{i-1} \Big[ \mb J^{(0)}_2(q_2,q_1), \mb C^{(ij)} ( q_{2\perp}^{(ij)}, \mu )\Big]
\tilde {\mb J}^{(0)} (q_1)
\, ,
\end{align}
yielding
\begin{align}
G_{2(a,b)}+ G_{2(c)} &= g^2 \Big(\sum_{i = 1}^{n+1} \sum_{j=1}^{i-1}
\mb C^{(ij)}(m, q_{2\perp}^{(ij)}) \mb J^{(0)}_{2} (q_2, q_1)+
\mb J^{(0)}_2(q_2,q_1) \mb C^{(ij)} ( q_{2\perp}^{(ij)}, \mu )\Big)
\tilde {\mb J}^{(0)} (q_1)
\nn \\
&= g^2 \Big( \sum_{i = 1}^{n+1} \sum_{j=1}^{i-1}
\mb C^{(ij)}(m, q_{2\perp}^{(ij)})\Big) \: \mb J^{(0)}_2 (q_2, q_1) \tilde {\mb J}^{(0)} (q_1)
\\
& +g^2 \mb J^{(0)}_2 (q_2, q_1) \Bigg[ \bigg( \sum_{i = 1}^n \sum_{j=1}^{i-1} \mb C^{(ij)} ( q_{2\perp}^{(ij)}, \mu )\Big)
\tilde {\mb J}^{(0)} (q_1)
+\sum_{j = 1}^n \sum_{\mathclap{\substack{k = 1\\ k\neq j}}}^n \mb C^{(q_1j)} (q_{2\perp}^{(q_1j)}, \mu) \, \mb d^{(0)}_{jk}(q_1) \Bigg]
\nn
\, ,
\end{align}
where we have rewritten the expression in anticipation of combining it with the Wilson coefficient in \eq{Cnp1} by splitting the second term in the first line into cases with $i \leq n$ and $i = n+1$. We then expressed $\tilde {\mb J}^{(0)}(q_1)$ for $i = n+1 = q_1$ in terms of $\mb d^{(0)}_{jk}(q_1)$ for each $j$ using \eq{J20Def}.

Finally, the rescattering graphs sum up as
\begin{align}\label{eq:G2c}
G_{2(d,e,f)} &= g^2\, \sum_{i = 1}^n \bigg[ \left(\sum_{\mathclap{\substack{j = 1\\ j\neq i}}}^n
\mb C^{(q_2 j )}(m , q_{2\perp}^{(ij)}) \, \mb d^{(0)}_{ji} (q_2)\right) \nn \\
& \quad+ \mb C^{(q_2 q_1 )}(m , q_{2\perp}^{(q_1i)}) \, \mb d^{(0)}_{q_1i} (q_2) + \mb C^{(q_2 i )}(m , q_{2\perp}^{(q_1 i )}) \, \mb d^{(0)}_{iq_1} (q_2)
\bigg] \tilde {\mb J}^{(0)} (q_1) \,, \nn\\
&= g^2\, \sum_{i=1}^{n+1} \sum_{\mathclap{\substack{j = 1\\ j\neq i}}}^{n+1} \mb C^{(q_2 i)}(m, q_{2\perp}^{(ij)}) \, \mb d^{(0)}_{ij}(q_2) \, \tilde {\mb J}^{(0)} (q_1) \, .
\end{align}
where the first line represents the cases where $q_2$ is sourced by one of the original $n$ hard partons and also rescatters against another one of them (\fig{1L2R}d). The second line accounts for cases in \fig{1L2R}e,f where the $q_2$ gluon is sourced by the $q_1$ gluon and rescatters against some other hard parton, and vice versa.

Having assembled all the ingredients, we are now in the position to write down the imaginary part of the one-loop, two-real emissions amplitude:
\begin{align}\label{eq:AMFS_SCET}
&\Im
\bigg( \langle( q_2, \veps_2) ,( q_1, \veps_1) ,\{p_i\} | O_{n+1}( \{ q_{1\perp}^{(ij)}, n_1, \omega_i ,n_i\}) | 0 \rangle \Big | {\cal C}_{n+1} (\{ q_{1\perp}^{(ij)}, \omega_i \},\mu) \Big]^{(1)} \bigg)\nn \\
&=
\bigg( g^2 \mb J^{(0)}_2 (q_2,q_1) \tilde{\mb J}^{(0)} (q_1) \, \Im \left|{\cal C}^{(1)}_{n+1} (\{\omega_i\},q_{1\perp}^{(ij)},\mu) \right]
+ G_{2(a+b+c)} (m , q_2, q_1, \mu) \big|M_0^{(0)}\big]
\bigg) \nn
\\
&=
g^2 \, \Bigg[ \mb J^{(0)}_2 (q_2,q_1) \tilde{\mb J}^{(0)}(q_1)\sum_{i = 1}^n \sum_{\mathclap{\substack{j = 1\\ j\neq i}}}^n
\mb C^{(ij)}(q_{1\perp}^{(ij)},\sqrt{\omega_{ij}}) \\
&\qquad \quad + \mb J^{(0)}_2(q_2,q_1) \sum_{i = 1}^{n}
\bigg(
\sum_{j=1}^{i-1}
\mb C^{(ij)} ( q_{2\perp}^{(ij)}, q_{1\perp}^{(ij)} )
\tilde{\mb J}^{(0)} (q_1)
+ \sum_{\mathclap{\substack{j = 1\\ j\neq i}}}^n \mb C^{(q_1i)} (q_{2\perp}^{(q_1i)}, q_{1\perp}^{(ij)}) \, \mb d^{(0)}_{ij}(q_1)
\bigg)
\nn \\
&\qquad \quad
+
\sum_{i = 1}^{n+1} \bigg( \sum_{j=1}^{i-1} \mb C^{(ij)}(m, q_{2\perp}^{(ij)}) \mb J^{(0)}_2 (q_2, q_1)
+ \sum_{\mathclap{\substack{j = 1\\ j\neq i}}}^{n+1} \mb C^{(q_2 i)}(m, q_{2\perp}^{(ij)})\, \mb d^{(0)}_{ij}(q_2) \bigg) \tilde{\mb J}^{(0)} (q_1)\Bigg] \Big| M_0^{(0)}\Big]\, .\nn
\end{align}
Thus we have derived the expected result: the first line in the final equation represents all the hard interactions above the scale of $q_{1\perp}$, the second being a straightforward generalization of the one-loop, one-emission result in \eq{ImMn} where the momenta $\{q_{2\perp}^{(ij)}\}$ are now seen as the infrared cutoff, or equivalently, the momenta $\{q_{1\perp}^{(ij)}\}$ are seen as the hard scales in the low energy EFT.
By repeated operations, we thus arrive at the AMFS result in \eq{AMFS}.

\section{Ordered limit of the double soft emission amplitude}
\label{sec:DoubleSoft}
In this section we make a correspondence with the calculations of \Ref{Angeles-Martinez:2015rna} by calculating imaginary part of the one-loop, two soft gluon emission amplitudes~\cite{Zhu:2020ftr} in the ordered limit.
The ordered limit of two soft emissions includes the soft limit we considered above where $q_2^\mu \ll q_1^\mu$, but also additional cases considered in \Ref{Angeles-Martinez:2015rna} where only the transverse momenta $q_{2\perp}^\mu \ll q_{1\perp}^\mu$ are required to be hierarchical and either of the two emissions is allowed to be collinear with another hard parton (with still Eikonal coupling to the hard partons). Thus, with the double emission amplitude, we can test configurations beyond those that can be described above in the recursive EFT.

The key strategy followed in \Ref{Angeles-Martinez:2015rna} was to carefully group the full theory graphs with Eikonal cuts. This, however, amounted to considering numerous different groupings, with the individual diagrams being very different in each region. The correct ordering variable was shown to emerge through a highly non-trivial interplay of many different orderings in many different individual diagrams. By repeating the double soft emission calculation in SCET we will demonstrate that not only one encounters a lot fewer diagrams, but also that the grouping necessary for the full theory derivation is already \textit{implicit} in the SCET graphs. As a result, every single diagram in the ordered limit has a unique and intuitive contribution to the AMFS result in \eq{AMFS_SCET}.

\begin{figure}[t]
\centering
\includegraphics[width=0.7\textwidth]{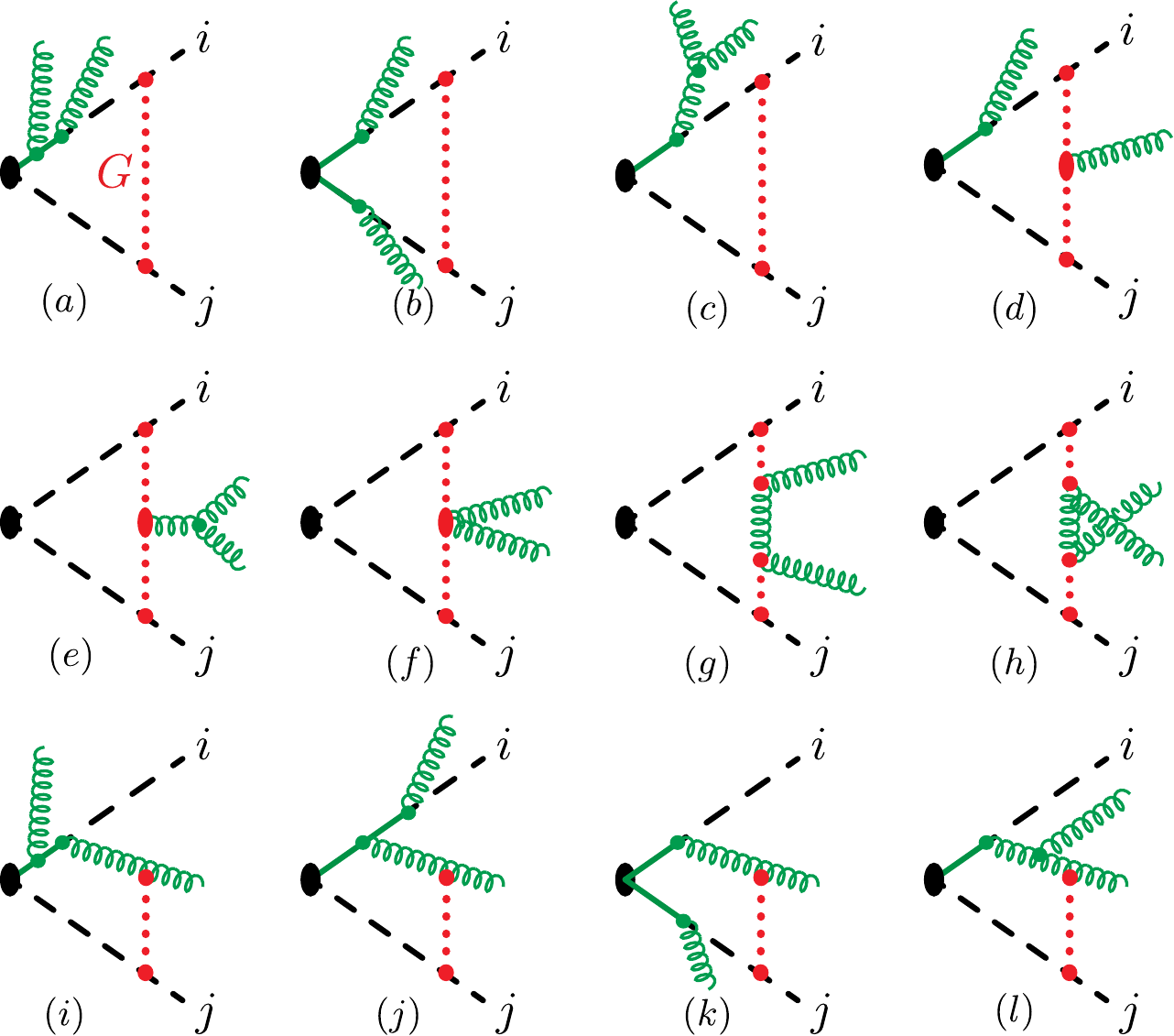}
\caption{One loop double soft emission graphs with a Glauber exchange. Additional diagrams are obtained by considering attachments to other collinear legs.}
\label{fig:2gall}
\end{figure}

The graphs for double soft emissions are shown in \fig{2gall}. For these one-loop graphs to yield an imaginary part there must be a Glauber exchange, which can be between any two of the $n$ collinear legs $i$ and $j$, or between one of the outgoing collinear partons $i$ and a soft emission. The two soft emissions can be sourced by either the Wilson lines in the hard scattering operator $O_n$ in \eq{OnSCET} or from the Wilson lines in the mid-rapidity operator $O_{n_isn_j}^{(ij)}$ in \eq{OGSCET} as a part of a forward scattering between the legs $i$ and $j$. Additional cases arise when we allow for the two soft emissions to be separated in rapidity, in addition to being ordered in momentum. In this scenario Glauber exchanges between the two gluons become possible, as well as the cases where the softer of the two gluon can be produced via Lipatov vertex between the other gluon and a collinear leg. Here, however, one implicitly assumes that one of the energetic soft gluon is resolved, and cannot be derived starting from the ordered limit of the double soft gluon emission amplitude.

For simplicity (and for technical reasons), we also do not consider the graphs \fig{2gall}i-l where the Glauber exchange happens between $q_1$ and another hard parton. As shown in \app{Rescatter}, these rescattering graphs are essentially given by their zero bins as the naive graphs evaluate to zero. Here, in the presence of another softer emission, one must, however, be careful with taking the Glauber limit of the soft propagators with $q_1$ gluon for the zero bin subtractions, as we also demand $q_2^\mu \ll q_1^\mu$, and the two limits do not commute. As a result these graphs become unwieldy very quickly, and they are best evaluated via the two-step EFT matching as described above in \sec{AddSofter}. They are nevertheless comparatively simpler to deal with in SCET than the corresponding full theory cut graphs.
Thus, we consider the following classes of diagrams
\begin{enumerate}
\item Double soft emission from the hard scattering operator and Glauber exchange between $i$ and $j$ hard partons (\fig{2gall}a-c)
\item One soft gluon from the hard scattering operator and other as a part of forward scattering between $i$ and $j$ hard partons (\fig{2gall}d).
\item Double soft emission from forward scattering of $i$ and $j$ hard partons via mid rapidity operator $O^{ij}_{n_isn_j}$ (\fig{2gall}e,f).
\item Double soft emission from forward scattering of $i$ and $j$ hard partons via T-product of $O^i_{n_is}$ and $O^j_{n_js}$ operators (\fig{2gall}g,h).
\item Double soft emission from the hard scattering operator and a Glauber exchange between the softer emission $q_2$ and another outgoing hard parton (\fig{2gall}i-l).
\end{enumerate}

The diagrams in the category 1, shown in \fig{2gall}a-c, involve tree level soft emission with a subsequent Glauber exchange between the hard partons. As shown in \Ref{Angeles-Martinez:2015rna}, the result for tree level double soft emission, $\mb K_2(q_1,q_2)$, reduces to that in \eq{tree2R}, not only when $q_1$ and $q_2$ are ordered and at wide angles to the hard parton, but also for cases where $q_1$ or $q_2$ are collinear to one of the hard partons, and only their transverse momenta are hierarchical, as mentioned above. Hence, we have
\begin{align}\label{eq:treeFactorise}
\mb K_{2}^{C_1 C_2} (q_1, q_2) \Big|_{q_2 \sim \rho q_1,\: \rho \ll 1} = \big[C_1 , C_2 \big | \mb J^{(0)}_{2}(q_2, q_1) \tilde {\mb J}^{(0)} (q_1) \, .
\end{align}
Hence, the result for graphs in \fig{2gall}a-c is straightforward:
\begin{align}\label{eq:G2ac}
S_2^{(a,b,c)} = g^2 \, \sum_{i=1}^n \sum_{j=1}^{i-1} \mb C^{(ij)}(m, \mu) \mb J^{(0)}_{2}(q_2, q_1) \tilde {\mb J}^{(0)} (q_1) \, ,
\end{align}

The category 2 shown in \fig{2gall}d involves combination of the graphs (a) and (b) in \fig{1L1R}. The soft gluon emission from the hard vertex simply decouples from the Glauber exchange. Using \eq{G1b} we have
\begin{align}\label{eq:G2d}
S_2^{(d)}= g^2 \sum_{i=1}^n\sum_{j=1}^{i-1} \bigg( \Big[ \tilde{\mb J}^{(0)} (q_1) ,\: \mb C^{(ij)} ( q_{1\perp}^{(ij)}, \mu )\Big] \tilde{\mb J}^{(0)} (q_2) + \Big[ \tilde{\mb J}^{(0)} (q_2) ,\: \mb C^{(ij)} ( q_{2\perp}^{(ij)}, \mu )\Big] \tilde {\mb J}^{(0)} (q_1) \bigg) \, ,
\end{align}
where $\tilde {\mb J}_\mu^{(0)}(q_2)$ is always dotted with $\veps^{*\mu}(q_2)$.

Next, we consider diagrams in category 3, shown in \fig{2gall}e,f, where the double soft emission occurs as a part of forward scattering between $i$ and $j$ collinear legs sourced by the mid rapidity operator $O_{n_isn_j}$. The results are derived in \app{midRapidity} and are given by
\begin{align}\label{eq:G2ef}
S_2^{(e)} &= g^2 \Big(\mb T_{q_1} \frac{ q_1 \cdot \veps^* (q_2)}{q_1 \cdot q_2}\Big)
\big[
\tilde {\mb J}^{(0)}(q_1) , \, \mb C^{(ij)} (q_{1\perp}^{(ij)}, \mu) \big] \, , \\
S_2^{(f)} &= g^2\Big[ \tilde{\mb J}^{(0)}(q_{2}) , \, \big[
\tilde {\mb J}^{(0)}(q_1) , \, \mb C^{(ij)} (q_{1\perp}^{(ij)}, \mu)\big]\Big] \, .\nn
\end{align}

As explained in \app{SoftProp}, the graphs in \fig{2gall}g,h that involve T-product of soft-collinear forward scattering Glauber operators, $O_{n_is}$ and $O_{n_js}$ are subleading in the ordered limit $q_2^\mu \ll q_1^\mu$. This can be seen by noting that the scale of the Glauber momentum running down the loop is $\ell_\perp \sim q_{1\perp}$ and the graphs do not receive $1/(n_i\cdot q_2)$ enhancement unlike other graphs.

Finally, the results in Eqs.~(\ref{eq:G2ac}),(\ref{eq:G2d}) and (\ref{eq:G2ef}) can be combined to yield a more intuitive result:
\begin{align}\label{eq:S2ah}
S_{2}^{(a-h)} &= \mb C^{(ij)} (m, q_{2\perp}^{(ij)})\: \mb J^{(0)}_2(q_2,q_1) \: \tilde {\mb J}^{(0)}(q_1) \\
&\quad +
\mb J^{(0)}_2(q_2,q_1) \: \mb C^{(ij)} (q_{2\perp}^{(ij)} , q_{1\perp}^{(ij)}) \: \tilde {\mb J}^{(0)}(q_1)\nn \\
&\quad + \mb J^{(0)}_2(q_2,q_1)\: \tilde {\mb J}^{(0)}(q_1) \:
\mb C^{(ij)} (q_{1\perp}^{(ij)} ,\mu)
\nn \, .
\end{align}
When combined with the Wilson coefficient in \eq{ImCn} the $\mu$ in the result above is replaced by $\omega_{ij}$ and we recover the terms expected from the AMFS result in \eq{AMFS}. Note that in deriving this result we had to consider only a handful of EFT diagrams, as opposed to the analysis in \Ref{Angeles-Martinez:2015rna} that needed careful grouping of several full theory graphs into 20 different structures.

Finally, we now turn to the diagrams in \fig{2gall}i-l which correspond to category 5 and are evaluated in \app{rescatter2}. The result for graphs with Glauber exchange between $q_2$ and any other hard partons is given by \eq{Sq2jFull}. Adding to it contributions where $q_2$ exchanges Glauber with $q_1$ in \fig{1L2R}d and where $q_2$ is sourced by $q_1$ in \fig{1L2R}e, we get
\begin{align}
S_2^{(i-l)(q_2j)} = g^2 \sum_{i = 1}^{n+1}\sum_{j\neq i}
\mb C^{(q_2 j)} (m, q_{2\perp}^{(ij)})\: \mb d^{(0)}_{ji}(q_2)\: \tilde{\mb J}^{(0)} (q_1)\, .
\end{align}
which in combination with the second line in \eq{S2ah} corresponds to the terms in the last line in \eq{AMFS_SCET}. Again, for these rescattering graphs, the SCET derivation of two gluon amplitude ends up being a lot more tractable.

\section{Conclusions and future directions}
\label{sec:conclusion}

In this paper we have rederived the AMFS result~\cite{Angeles-Martinez:2015rna,Angeles-Martinez:2016dph} using Glauber SCET operators. We have shown that the result for $N$ soft gluon emission amplitude can be derived very efficiently by considering a sequence of EFTs where each time a new soft emission is resolved to become a collinear direction. The combination of the Wilson coefficient and handful of EFT diagrams significantly simplify an analysis that otherwise involves several QCD graphs where the intermediate stages bare little resemblance with the final result.

Furthermore, the Markovian nature of the AMFS result in \eq{AMFS} suggests that there is a deeper underlying physics that is obscured in the full theory analysis. By repeating the calculation in SCET every individual diagram directly contributes to a specific term in the AMFS result. We see a special role played by the Lipatov vertex which serves to implement an infrared cutoff on the virtual transverse momentum integral. Additionally, the EFT derivation clarifies why the transverse momentum scale of the soft gluon evaluated in the parent dipole frame must appear in the upper limit of the loop integrals. We saw that this is because for the \eftnp where the soft gluon is resolved, the transverse momentum in parent dipole frame emerges as the appropriate hard scale.
\begin{figure}[t]
\centering
\includegraphics[width=0.8\textwidth]{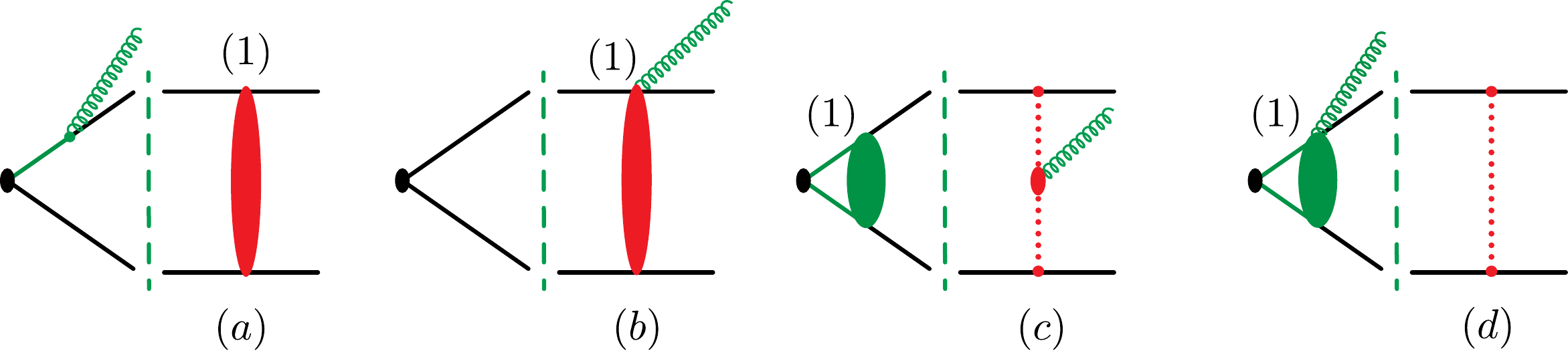}
\caption{Diagrams needed for extension of AMFS result to two-loops. Additional diagrams not shown include one-loop corrections to the $n$-$s$ forward scattering and soft emission diagrams involving 3 Wilson lines.}
\label{fig:extension}
\end{figure}

The SCET derivation thus has made it possible for us to envisage a tractable way forward in extending the AMFS result to higher orders. In \fig{extension} we show as an illustration some of the types of diagrams needed to extend the result for the imaginary part to next order. The single Glauber exchange is needed to obtain the imaginary part, and one thus needs an additional loop correction, either to the hard vertex or in the forward scattering process, namely the effective one-loop collinear-Glauber vertex, soft-Glauber vertex and the Lipatov vertex. Additionally, given the relatively simple nature of two emission graphs considered in \sec{DoubleSoft} one can generalize the result by breaking the strong ordering chain and letting the two soft emissions have commensurate energies. Finally, it will also be interesting to extend this analysis along the lines of \Ref{Schwartz:2017nmr} to include collinear emissions and examine the ordering variable. This further improve our understanding the resummation of superleading logarithms.

\subsubsection*{Acknowledgements}
I am grateful to Jeffrey Forshaw for introducing me to this problem and for numerous helpful discussions throughout the project. I thank Jack Holguin for helpful discussions, especially with the color space notation. I am also thankful to Thomas Becher for a very careful reading of the draft and providing many helpful comments and suggestions, and Iain Stewart for discussions and cross checking the two-gluon emission vertex Feynman rule. I am a member of the Lancaster-Manchester-Sheffield Consortium for Fundamental Physics, which is supported by the UK Science and Technology Facilities Council (STFC) under grant number ST/T001038/1.

\appendix

\section{Color space notation}
\label{app:Color}
In this appendix we clarify in detail the color space notation. The convention here bears some similarities to \Ref{Becher:2009kw}.
\subsection{Hard scattering operators}
A generic hard scattering operator for a process involving $n$ collinear fields can be expressed as
\begin{align}\label{eq:On}
O_n = \sum_\Gamma \int \bigg(\prod_{i = 1}^n \df \omega_i\bigg) {\cal C}_{n, \Gamma, \{a,\alpha\} } (\{\omega_i\}, \mu) \,
O_{n }^{\{a,\alpha\}} \big(\{\omega_i, n_i\}, \mu\big) \, ,
\end{align}
where the indices $\{a\}$ denote all the (adjoint and (anti-)fundamental) color indices, $\{\alpha\}$ corresponds to spin and Lorentz-vector indices, and $\Gamma$ runs over the possible Dirac structures.
The Wilson coefficient ${\cal C}_{n, \Gamma,\{a,\alpha\}}(\{\omega_i\}, \mu)$ depends only on the large momenta $\{\omega_i\}$ in each collinear sector. We have included the factors of Dirac matrices $\Gamma^{\alpha}$, spinors, and external Lorentz vectors corresponding to the $n$ hard partons in $ {\cal C}_{n, \Gamma, \{a,\alpha\} }$. For the present discussion, it is not necessary to decompose the color structures into a set of specific basis vectors. We will also suppress the $\mu$ dependence for simplicity. The operator $O_{n}^{\{a,\alpha\}}$ consists of both collinear fields and soft Wilson lines:
\begin{align}\label{eq:OmExplicit}
O_{n}^{\{a,\alpha\}} \big( \{\omega_i, n_i\} \big) &= \sum_{\{b'\} }\Big[O_{n}^{(0)} ( \{\omega_i, n_i\} )\Big]^{\{\alpha\}}_{\{b'\}} \Big[O^s (\{n_i\})\Big]^{\{b'\}, \{a\}} \, ,
\end{align}
where $O_{n}^{(0)}$ consists of all the quark and gluon collinear fields,
\begin{align}
\Big[O^{(0)}_{n}\big(x,\{\omega_i,n_i\}\big) \Big]^{\{\alpha\}}_{\{b\}} =\prod_{i} \big(\phi_{n_i,\omega_i} (x)\big)^{\alpha_i}_{b_i} \, .
\end{align}
The field $\phi$ generically stands for the following gauge invariant, collinear building blocks~\cite{Bauer:2001ct} with their respective color charges:
\begin{align}\label{eq:buildingBlocks}
&\text{outgoing quarks or incoming anti-quarks:}&
&\big(\phi_{n_i,\omega_i} \big)^{\alpha_i}_{b_i}\ra \big(\overline \chi_{n_i, \omega_i}\big)^{\alpha_i}_{b_i}\,,&
& ( T_i^c)_{ba} = t_{ba}^c\, ,&
\\
&\text{incoming quarks or outgoing anti quarks:}&
&\big(\phi_{n_i,\omega_i} \big)^{\alpha_i}_{b_i} \ra \big(\chi_{n_i, \omega_i}\big)^{\alpha_i}_{b_i} \,,&
& ( T_i^c)_{ba} = (\bar t)^c_{ba} \, ,&
\nn \\
&\text{gluons:}&
&\big(\phi_{n_i,\omega_i} \big)^{\alpha_i}_{b_i} \ra {\cal B}^{\mu_i B_i}_{n_i \perp, \omega_i} \,,&
& ( T_i^C)_{BA} = \im f^{BCA}\, ,& \nn
\end{align}
with analogous relations for the QCD fields.
We will also use $T_g$ to refer to matrices in adjoint representation. In general, unless explicitly stated, the representation associated with the collinear gluon field ${\cal B}_{n_i\perp}^{\mu A}$ will be adjoint.

At leading power, the soft gluons will always be produced from the soft Wilson lines contained in the operator $O_s(\{n_i\})$:
\begin{align}\label{eq:OsExplicit}
\big[O^s(\{n_i\})\big]^{\{b'\},\{a\}} &\equiv\prod_{i = 1}^{n_{\overline \chi}} \big[S^\dagger_{n_i\pm}\big]_{b_i a_i} \prod_{j = 1}^{n_{\chi}} \big[S_{n_j\pm}\big]_{a_j b_j} \prod_{k = 1}^{n_g} \big[{\cal S}_{n_k}\big]_{A_k B_k} \, ,
\end{align}
where $n_{\chi}$ ($n_{\overline \chi}$) is the number of collinear quark/anti-quark fields (field conjugates), and $n_g$ is the number of gluon fields present in the operator $O_{n,\Gamma}$. The subscript $+$ and $-$ denotes attachments to quarks and anti-quarks, respectively. One can further distinguish between direction of the Wilson lines as incoming or outgoing. For incoming and outgoing collinear quark fields, we have respectively
\begin{align}\label{eq:WilsonLinesExplicit}
S_{n_i+} = {\rm P} \exp \Big(-\im g \int_{-\infty}^0 \df s \: n_i \cdot A_s (x^\mu +\frac{n^\mu}{2}s )\Big) \,, \quad
S^\dagger_{n_i+} = {\rm P} \exp \Big(- \im g \int_0^\infty \df s \: n_i \cdot A_s (x^\mu +\frac{n^\mu}{2}s) \Big) \, .
\end{align}
Likewise, for incoming and outgoing collinear anti-quark fields, we have respectively
\begin{align}\label{eq:WilsonLinesExplicit2}
S^\dagger_{n_i-} = \overline {\rm P} \exp \Big(+ \im g \int_{-\infty}^0 \df s \: n_i \cdot A_s (x^\mu +\frac{n^\mu}{2}s) \Big) \, , \quad
S_{n_i-} = \overline {\rm P} \exp \Big(+\im g \int_0^\infty \df s \: n_i \cdot A_s (x^\mu +\frac{n^\mu}{2}s )\Big)
\, .
\end{align}
However, as noted in \Ref{Rothstein:2016bsq}, with inclusion of Glauber modes, the direction of soft Wilson lines in \eq{WilsonLines} in the Glauber and hard scattering operators becomes irrelevant. This is because the two cases of integration from $-\infty \ra 0$ and $0\ra \infty$ of in \eq{WilsonLinesExplicit} result in $n\cdot k \pm \im 0$ factors in the propagators (for incoming $k^\mu$). These poles are however already captured by the Glauber region, and any non-zero contribution is removed from the soft graphs upon Glauber zero-bin subtraction. Thus, we will simply take the direction to be outgoing for the diagrams considered in this paper.
Finally, the Wilson lines ${\cal S}_{n_k}$ corresponding to gluon fields are expressed in the adjoint representation. Note also that the sign in the exponential differs from the expressions of Wilson lines in SCET literature (see \Ref{Bauer:2001yt}, for example). This results from our choice of sign of $g$ typical in the QCD literature, for example \Ref{CATANI2000435}, which is negative of that in the SCET literature. Otherwise, using SCET convention for $g$ will result in negative of soft gluon current.

We can efficiently deal with the color by expressing \eq{On} as a dot product of vectors in color space:
\begin{align}\label{eq:OnVec}
&\big | {\cal C}_{n,\Gamma} \big] \equiv \sum_{\{a\}}{\cal C}_{n, \Gamma, \{a\} } \big|\{a\}\big]\,,&
&{\cal C}_{n, \Gamma, \{a\} } (\{\omega_i\} ) = \big [ \{a\} \big | {\cal C}_{n,\Gamma} (\{\omega_i\} )\big] & \\
&\big [ O_{n} \big | \equiv \sum_{\{a\}} O_{n}^{\{a\}} \big [ \{a\}\big| \, ,&
& O_{n}^{\{a\}} \big(\{\omega_i,n_i\}\big) = \big [ O_{n} \big(\{\omega_i,n_i\}\big)\big | \{a\} \big] \, ,&\nn
\end{align}
which leads to the compact notation in \eq{OnCompact}.
We reserve angle brackets for physical state vectors. For simplicity, we will take the basis $\big|\{a\}\big]$ to be orthonormal. More practical choices involve using over-complete basis of states, in which case the conjugate vectors will have to be carefully defined.

To deal with additional color matrices resulting from real and virtual radiative corrections, we will make use of color operators $\mb T_i^C$ that combine with the collinear fields in the following way:
\begin{align}\label{eq:TiAction}
\big(\chi_{n_i }\big)_a \mb T_j^C&= - (T^C)_{ab} \big( \chi_{n_i} \big)_b\delta_{ij} = -\big(T^C \chi_{n_i} \big)_a \delta_{ij} \,, \\
\big(\overline \chi_{n_i} \big)_a \mb T_j^C &= \big(\overline \chi_{n_i} \big)_b\big(T^C)_{ba} \delta_{ij} = \big(\chi_{n_i} T^C)_{a} \delta_{ij} \, ,\nn \\
{\cal B}^{\mu_i A}_{n_i \perp, \omega_i}\, \mb T_j^C &= \im f^{ACB} {\cal B}^{\mu_i B}_{n_i \perp, \omega_i} \delta_{ij}\, . \nn
\end{align}
Although it is conventional to let the operator $\mb T_j$ act on fields from the left, as we did above for the full theory amplitude in \eq{Ma1}, the notation here is consistent with expressing the operator $O_{n,\Gamma}^{\{a\}}$ as a bra-vector in the color space in \eq{OnVec}, and additionally, along with expressing Wilson coefficients as ket-vectors, it leads to natural ordering of color operators when evaluating (real or virtual) matrix elements. The action of the color operator on basis vectors $\big|\{a\}\big]$ is given by
\begin{align}\label{eq:TiVec}
\mb T_i^C \big | \ldots , a_i , \ldots \big ] &=
\sum_{b_i'} \big | \ldots , b_i' , \ldots \big ] \big [ \ldots , b_i' , \ldots \big | \mb T_i^C \big | \ldots , a_i , \ldots \big ]
\\
&=
\sum_{b_i'} \big | \ldots , b_i' , \ldots \big ] \big(\mb T_i^C\big)_{b_i' a_i} \nn \, ,
\end{align}
where in the first line we used the completeness relation. Here we will only be concerned with external soft gluon emissions, so that the upper adjoint index $C$ on the operator $\mb T_i^C$ will always be associated with an additional soft real emission, whereas the lower indices will be related to the matrix operation of the operator on the collinear fields $\phi_{n_i}$. For example,
\begin{align}
\big [ C \big | \mb T_i \equiv \mb T_i^C \, , \qquad
\big [ C_2 C_1 \big | \mb T_{1} \big | A_1 \big ]
=\big( \mb T_1^{C_2}\big)_{C_1 A_1} = \im f^{C_1 C_2 A_1} \, ,
\end{align}
where $\mb T_1$ is the color generator associated with the soft gluon with index 1. The $\mb T_i$ operator thus, by adding an emission, increments the dimension of the space.

In terms of color operators, the Wilson lines for various cases \eq{OsExplicit} (for outgoing direction) can be combined into a single formula:
\begin{align}\label{eq:WilsonLines}
\mb S_{n_i} &\equiv { \mb P }\exp \bigg(-\im g \int_0^\infty \df s\: n_i\cdot A_s^A\Big (y^\mu + s \frac{n^\mu}{2}\Big) \mb T_i^A\bigg)
\, .
\end{align}
The representation, unless explicitly indicated, will be taken to be the one associated with the corresponding $i^{\rm th}$ collinear field.

In terms of \eq{WilsonLines} the vector $\big[O_{n} \big|$ can be expressed as
\begin{align}
\big[ O_{n}\big(\{\omega_i,n_i\}\big)\big | \equiv \big[ O_{n}^{(0)}\big(\{\omega_i,n_i\}\big) \big| \prod_{i = 1}^{n} \mb S_{n_i} \, .
\end{align}
Using the relation in \eq{TiVec} we can rewrite the components of $\big[O_{n} \big|$ shown in \eq{OmExplicit} as
\begin{align}
\big[ O_{n}\big(x,\{\omega_i,n_i\}\big)\big | \{a\} \big ]
&= \sum_{\{b\}}
\big[ O_{n}^{(0)}\big(x,\{\omega_i,n_i\}\big)\big | \{b\} \big ]
\big [ \{b\} \big | \Big(\prod_{i = 1}^{n} \mb S_{n_i} \Big)\big | \{a\} \big ]
\, .
\end{align}
As noted above in \sec{EFTsetup}, the matrix elements of the operators, unless tree level, will turn the vector $[O_n|$ into a matrix in color space due to color mixing at higher orders.
\subsection{Glauber operators}
\label{app:GlauberOps}
The Glauber action $S_G$ is composed of Glauber potential operators that mediate interactions between two collinear sectors (with an intermediate soft sector) and between a collinear and a soft sector:
\begin{align}\label{eq:SG}
S_G = \int \df^4 x \: e^{-\im x \cdot {\cal P}} \Bigg[ \sum_{n_i, n_j} \sum_{i,j = q,g} O_{n_i s n_j}^{ij} ( x)+ \sum_{n_i} \sum_{i,j = q,g} O_{n_is}^{ij} ( x)\Bigg]
\end{align}
Here ${\cal P}^\mu$ is the label momentum operator that selects ${\cal O}(1)$ and ${\cal O}(\lambda)$ momenta, and is discussed further in \app{FeynRules}. The coordinate $x$ is conjugate to subleading momentum components.
Since the Glauber operators do not get hard physics corrections and do not get renormalized, we find it simplest to not decompose the operators in color space basis as above. In the color operator notation, the mid-rapidity Glauber operator for two collinear sectors $n_i$ and $n_j$ and the soft sector, is given by
\begin{align}\label{eq:Onsn}
O^{ij}_{n_i s n_j} &= \mb O_{n_i}^{i} \cdot \frac{1}{{\cal P}_\perp^2} \hat { \mb O}^{(n_in_j)}_s \frac{1}{{\cal P}_\perp^2}\cdot \mb O_{n_j}^{j} \, ,
\qquad
O^{ij}_{n_is} = \mb O_{n_i}^{i} \cdot \frac{1}{{\cal P}_\perp^2} \mb O_s^{n_i,j}
\end{align}
where the collinear operators are given by
\begin{align}\label{eq:CollinearBilinears}
\mb O_{n_i}^{q} = \overline \chi_{n_i} \mb T_i \frac{\bnslash_i}{2} \chi_{n_i} \, , \qquad
\mb O_{n_i}^g = \frac{1}{2} {\cal B}_{n\perp\mu}^A \mb T_g \frac{\bn_i}{2} \cdot ({\cal P} + {\cal P}^\dagger) {\cal B}_{n\perp}^{A\mu} \, ,
\end{align}
where $\mb T_g$ corresponds to the adjoint representation.
The single rapidity soft operators $\mb O_s^i$ in this notation read
\begin{align}\label{eq:SoftBilinears}
\mb O^{n_i, q}_s = 8\pi \alpha_s \Big(\overline \psi_S^{n_i} \mb T_i \frac{\nslash}{2} \psi_S^{n_i}\Big) \, , \qquad
\mb O^{n_i, g}_s=8\pi \alpha_s \Big( \frac{1}{2} {\cal B}_{S\perp\mu}^{n_iA} \mb T_g \frac{n_i}{2} \cdot ({\cal P} + {\cal P}^\dagger) {\cal B}_{S\perp}^{n_iA\mu}\Big) \, .
\end{align}
The mid rapidity soft operator is given by\footnote{The difference in the sign relative to \Ref{Rothstein:2016bsq} result from setting $g\ra -g$. See the note below \eq{WilsonLinesExplicit}.}
\begin{align}\label{eq:MidRapidityOp}
\hat { \mb O}^{(n_in_j)}_s = 8\pi \alpha_s &\Big[ {\cal P}_\perp^\mu \mb S_{n_i}^{\mb g \dagger} \mb S^{\mb g }_{n_j} {\cal P}_{\perp\mu}
+{\cal P}_\perp^\mu g\mb B_{S\perp\mu}^{\mb g n_i} \mb S_{n_i}^{\mb g \dagger} \mb S^{\mb g }_{n_j}
+ \mb S_{n_i}^{\mb g \dagger} \mb S^{\mb g }_{n_j} g \mb B_{S\perp}^{\mb g n_j\mu} {\cal P}_{\perp\mu}
+ g\mb B_{S\perp}^{\mb g n_i\mu} \mb S_{n_i}^{\mb g \dagger} \mb S_{n_j}^{\mb g } g\mb B_{S\perp\mu}^{\mb g n_j} \nn \\
&+ \frac{n_{i\mu} n_{j\nu}}{2} \mb S_{n_i}^{\mb g \dagger} \im g \mb G_S^{\mb g \mu\nu} \mb S^{\mb g }_{n_j}
\Big] \, ,
\end{align}
where the superscript $\mb g$ emphasizes that the representation associated with the objects in $\hat {\mb O}_s$ is adjoint, unrelated to the representation associated with the $i$ and $j$ collinear sectors. The fields $ \mb B_{S\perp}^{\mb g n_j\mu}$ and $\mb G_S^{\mb g \mu\nu}$ are defined as
\begin{align}
\mb B_{S\perp}^{\mb g n_j \mu } \equiv {\cal B}_{S\perp}^{C n_j \mu } \mb T_g^C \, ,
\qquad
\mb G_S^{\mb g \mu\nu} \equiv G^{\mu\nu C}_S \mb T_g^C \, .
\end{align}
We warn the reader that unlike \eqs{TiAction}{SoftBilinears} this equation is not to be interpreted as action of $\mb T_g^C$ on the soft gluon operators.
The dot product in \eq{Onsn} in fact contracts the adjoint indices that are superscripts on $\mb T_i$ in $\mb O_{n_i}$ and matrix entries of $\hat {\mb O}^{(n_in_j)}_s$. For example,
\begin{align}
\mb O_{n_i}^q \cdot\hat { \mb O}^{(n_in_j)}_s \cdot \mb O_{n_j}^q &\equiv \mb O_{n_i}^q | A ] [ A | \hat { \mb O}^{(n_in_j)}_s| B ] [ B | \mb O_{n_j}^q \nn \\
&= \Big( \overline \chi_{n_i} \mb T_i^A \frac{\bnslash_i}{2} \chi_{n_i}\Big) [ A | \hat { \mb O}^{(n_in_j)}_s| B ]
\Big( \overline \chi_{n_j} \mb T_j^B \frac{\bnslash_j}{2} \chi_{n_j}\Big)
\end{align}

\section{Feynman rules}
\label{app:FeynRules}

We now sketch derivation of Feynman rules for Glauber operators involving soft fields described in \Ref{Rothstein:2016bsq}.

\subsection{Momentum flow}
In SCET, the fields are distinguished by their momentum scaling, and thus it is convenient to make their large momentum components explicit:
\begin{align}\label{eq:phiMom}
\phi(\tilde x) = {\sum_{p_\ell , k_s}}^\prime \int \dfbar^2 q_\perp \phi_{p^{\pm}_\ell , k^{\pm}_s} (\tilde x, q_\perp) \, , \quad p^{\pm}_\ell \sim {\cal O}(1) , \quad k^{\pm}_s , q_\perp^\mu \sim {\cal O}(\lambda) \, .
\end{align}
Here $\dfbar k \equiv \df k/(2\pi)$ and the coordinates $\tilde x$ only involve the $\pm$ light-cone components following the decomposition in \eq{LCDef}:
\begin{align}
\tilde x^\mu \equiv \frac{n_i^\mu}{2} x^- + \frac{\bar n_i^\mu}{2} x^+ \, .
\end{align}
The `$\prime$' on the sum corresponds to avoiding the zero bins. For $n_i$-collinear fields this corresponds to $\bn_i\cdot p_\ell \neq 0$ and for soft fields $k_s^\pm \neq 0$.
Since we are interested in distinguishing fields with different scaling of momentum components, we work with the label momentum operator ${\cal P}^\mu$ that selects the ${\cal O}(1)$ and ${\cal O}(\lambda)$ momentum components:
\begin{align}
{\cal P}^\mu = \frac{n_i^\mu}{2} \big({\cal P}^- + {\cal P}_s^-\big) + \frac{\bn_i^\mu}{2}\big({\cal P}^+ + {\cal P}_s^{+}\big)+ {\cal P}_\perp^\mu \, , \qquad {\cal P}^\pm \sim \lambda^0 \, , \qquad {\cal P}_s^\pm \sim {\cal P}_\perp^\mu \sim \lambda \, .
\end{align}
The action of label operator on collinear and soft fields and field conjugates is given by
\begin{align}
{\cal P}^{\pm} \phi_{p_\ell} (x) = p_{\ell }^{\pm} \phi_{p_\ell}(x) \, , \qquad {\cal P}^{\pm} \phi^\dagger_{p_\ell} (x) = - p_{\ell }^{\pm} \phi^\dagger_{p_\ell}(x) \, .
\end{align}
Likewise, the conjuagate operator ${\cal P}^{\dagger\mu}$ acts on the fields on the left:
\begin{align}
\phi_{p_\ell}^\dagger (x) {\cal P}^{\dagger\pm} = p_{\ell }^{\pm} \phi^\dagger_{p_\ell}(x) \, , \qquad \phi_{p_\ell} (x){\cal P}^{\dagger\pm} = - p_{\ell }^{\pm} \phi_{p_\ell}(x) \, .
\end{align}
Similarly, the operators ${\cal P}_s^{\pm}$ pick out ${\cal O}(\lambda)$ sub-label momentum:
\begin{align}
{\cal P}_s^{\pm} \phi_{p_\ell, k_s} (x) = k_s^{\pm} \phi_{p_\ell, k_s} (x) \,, \qquad {\cal P}_{\perp}^\mu \phi_{p_\ell, k_s} (\tilde x, q_\perp) = q_\perp^\mu \phi_{p_\ell, k_s} (\tilde x, q_\perp) \, .
\end{align}
The action of the label operator on an operator consisting of a string of fields and field conjugates simply amounts to adding all the label momentum eigenvalues:
\begin{align}
{\cal P}^\mu \phi^\dagger_{p_{\ell_1}} \phi^\dagger_{p_{\ell_2}} \ldots \phi_{k_{\ell_1}} \phi_{k_{\ell_2}} \ldots &=\big( - p_{\ell_1}^\mu - p_{\ell_2}^\mu - \ldots + k_{\ell_1}^\mu + k_{\ell_2}^\mu + \ldots\big)\phi^\dagger_{p_{\ell_1}} \phi^\dagger_{p_{\ell_2}} \ldots \phi_{k_{\ell_1}} \phi_{k_{\ell_2}} \ldots \, , \nn \\
\phi^\dagger_{p_{\ell_1}} \phi^\dagger_{p_{\ell_2}} \ldots \phi_{k_{\ell_1}} \phi_{k_{\ell_2}} \ldots {\cal P}^{\mu\dagger}&= \big(p_{\ell_1}^\mu + p_{\ell_2}^\mu + \ldots - k_{\ell_1}^\mu - k_{\ell_2}^\mu - \ldots \big)\phi^\dagger_{p_{\ell_1}} \phi^\dagger_{p_{\ell_2}} \ldots \phi_{k_{\ell_1}} \phi_{k_{\ell_2}} \ldots\, .
\end{align}
From \eq{SG} we see that the $e^{i x \cdot {\cal P}}$ factor accounts for the Fourier phases involving ${\cal O}(1)$ and ${\cal O}(\lambda)$ momentum components, and the $\tilde x$ dependence thus corresponds to long distance $\sim 1/\lambda^2$ fluctuations.

We can likewise decompose the collinear and soft bilinears in \eqs{CollinearBilinears}{SoftBilinears} as sum over operators injecting definite label and sublabel momenta. We first consider the $O_{n_i s}^{ij}$ operator in \eq{Onsn}:
\begin{align}
\mb O_{n_i}^{i} (\tilde x) &= \sum_{ k^-_s} \int \dfbar^2 q_{\perp} \frac{\dfbar p_r^+\: \dfbar p_r^-}{2} e^{\im \frac{x^+ \, p_r^-}{2} + \im \frac{x^- \,p_r^+}{2}} \big[\mb O_{n_i, k_s^- }^{i}( p_r^{\pm} , q_\perp)\big] \, ,
\\
\mb O_s^{n_i, j} (\tilde x) &= \sum_{ k^-_s} \int \dfbar^2 q_{\perp} \frac{\dfbar p_r^+\: \dfbar p_r^-}{2} e^{\im \frac{x^+ \, p_r^-}{2} + \im \frac{x^- \,p_r^+}{2}} \big[\mb O_{s,k_s^-}^{n_i, j}( p_r^{\pm} , q_\perp)\big] \nn \, ,
\end{align}
where the operator $\mb O_{n_i,k_s^-}^i(p_r^\pm, q_\perp)$ is defined via
\begin{align}
\mb O_{n_i,k_s^-}^{i} (\tilde x, q_\perp )
&\equiv \big[\mb O_{n_i}^{i}(\tilde x ) (2\pi)^2\delta^2 (q_\perp - {\cal P}^\dagger_\perp) \delta_{{\cal P}_s^{-\dagger}, k_s^-}\big] \, , \\
&=
\int \frac{\dfbar p_r^+\: \dfbar p_r^-}{2} e^{\im \frac{x^+ \, p_r^-}{2} + \im \frac{x^- \,p_r^+}{2}} \big[\mb O_{n_i, k_s^- }^{i}( p_r^{\pm} , q_\perp)\big] \, . \nn
\end{align}
We see that because of ${\cal P}_\perp^\dagger$ and ${\cal P}_s^{-\dagger}$, the operator \textit{injects} a definite amount of ${\cal O}(\lambda)$ $\perp$ and $-$ sublabel momenta.
The quark operator in \eq{CollinearBilinears}, for example, is given by
\begin{align}\label{eq:QuarkExplicit}
\mb O_{n_i,k_s^-}^{q}(\tilde x, q_\perp) \equiv \sum_{k_s^{\prime-}} \int \dfbar^2 p_\perp \: \overline \chi_{n,k_s^{\prime-} + k_s^-} (\tilde x, p_\perp + q_\perp) \mb T_i \frac{\bnslash_i}{2} \chi_{n,k_s^{\prime -}} (\tilde x, p_\perp) \, ,
\end{align}
and similarly for the soft operator $\mb O_s^{n_i, q}$. From this expression, we first note that the collinear bilinears in \eq{CollinearBilinears} necessarily have zero large label ${\cal O}(1)$ momentum eigenvalue to maintain forward scattering kinematics:
\begin{align}
\mb O_{n_i}^{i} {\cal P}^{-\dagger} = 0 \, .
\end{align}
However, the operator is allowed to inject ${\cal O}(\lambda)$ sub-label momenta as indicated by $\bn_i \cdot k_s$ subscript and $q_\perp$ argument in \eq{QuarkExplicit}. Finally, implementing momentum conservation between soft and collinear operators, we get
\begin{align}\label{eq:OnsFull}
\int \df^4 x \: e^{\im {\cal P} \cdot x} \: O_{n_i s}^{ij} (x) = \int \frac{\df x^+ \df x^-}{2} \sum_{k_s^-} \int \frac{\dfbar^2 q_\perp}{q_\perp^2} \: \mb O_{n_i,k_s^-}^{i} (\tilde x, q_\perp ) \cdot \mb O_{s,- k_s^-}^{n_i,j}(\tilde x, -q_\perp) \, .
\end{align}
Here we see that $1/{\cal P}_\perp^2$ operator in \eq{Onsn} acts on the right and picks up the $\perp$-momentum transferred into the collinear sector on the left. We also note that to preserve the $Q(\lambda^2, 1 ,\lambda)$ power counting of the $n_i$-collinear sector, the operator does not inject any $ k_s^+ = n_i \cdot k_s \sim \lambda $ momentum. As a result the $n_i\cdot k_s$ component of the soft gluon remains unchanged in the scattering described by $O_{ns}$ operators.

We now turn to the mid-rapidity operator $\hat { \mb O}^{(n_in_j)}$ in \eq{MidRapidityOp}. Here we find cases with ${\cal P}_\perp^\mu$ inserted to the left or right of product of soft operators. The rule simply says that the ${\cal P}_\perp^\mu$ on the left (as in ${\cal P}_\perp^\mu \mb B_{S\perp\mu}^{\mb g n_i} \mb S_{n_i}^{\mb g \dagger} \mb S^{\mb g }_{n_j} $ ) picks up the $\perp$-momentum \textit{ejected from} everything to the right, and hence always includes the $\perp$-momentum transferred from the $n_j$-collinear sector. On the other hand, the ${\cal P}_{\perp\mu}$ operator to the right (as in $\mb S_{n_i}^{\mb g \dagger} \mb S^{\mb g }_{n_j} \mb B_{S\perp}^{\mb g n_j\mu} {\cal P}_{\perp\mu}$) will only pick up the transverse momentum transferred from the $n_j$-collinear sector. In other words,
\begin{align}\label{eq:OnsnMom}
\int \df^4 x \: e^{\im {\cal P}\cdot x} O_{n_i s n_j}^{ij} (x) &= \int \frac{\df x^+ \df x^-}{2} \sum_{\bar n_i \cdot k_s , \bar n_j \cdot k_s}
\int \frac{\dfbar^2 q_\perp}{q_\perp^2} \frac{\dfbar^2 q_\perp'}{q_\perp^{\prime\,2}} \\
&\qquad \times \mb O_{n_i, \bar n_i \cdot k_s}^i (\tilde x, q_\perp) \cdot \hat {\mb O}_{s,- \bar n_i \cdot k_s, -\bar n_j \cdot k_s}^{(n_in_j)} (\tilde x, -q_\perp, q_\perp^\prime) \cdot \mb O_{n_j,\bar n_j \cdot k_s}^{j} (\tilde x ,- q_\perp^{\prime}) \nn \, ,
\end{align}
where
\begin{align}\label{eq:MidRapidityOp2}
\hat {\mb O}_{s}^{(n_in_j)} (\tilde x, -q_\perp, q_\perp^\prime)
&= 8\pi \alpha_s (2\pi)^2 \delta^2 (q_\perp -q_\perp^\prime -{\cal P}_\perp) \Big[ q_\perp \cdot q_\perp' \mb S_{n_i}^{\mb g \dagger} \mb S^{\mb g }_{n_j}
+q_\perp^\mu g\mb B_{S\perp\mu}^{\mb g n_i} \mb S_{n_i}^{\mb g \dagger} \mb S^{\mb g }_{n_j}
\\
&+ \mb S_{n_i}^{\mb g \dagger} \mb S^{\mb g }_{n_j} g \mb B_{S\perp}^{\mb g n_j\mu} q^\prime_{\perp\mu}
+ g\mb B_{S\perp}^{\mb g n_i\mu} \mb S_{n_i}^{\mb g \dagger} \mb S_{n_j}^{\mb g } g\mb B_{S\perp\mu}^{\mb g n_j}
+ \frac{n_{i\mu} n_{j\nu}}{2} \mb S_{n_i}^{\mb g \dagger} \im g \mb G_S^{\mb g \mu\nu} \mb S^{\mb g }_{n_j}
\Big]\, .\nn
\end{align}
Thus, from \eq{OnsnMom}, the total outgoing soft momentum is given by
\begin{align}
k_s^\mu = -\Big(\frac{n_i^\mu}{2} \bn_i\cdot k_s + \frac{n_j^\mu}{2} \bn_j\cdot k_s + q_\perp^\mu - q_\perp^{\prime\mu} \Big)\, .
\end{align}

\subsection{Feynman rules for Glauber operators}
\label{app:FeynGlauber}
In this section we describe the general strategy for deriving the Feynman rules for Glauber operators stated in \Ref{Rothstein:2016bsq}. We will restrict the discussion to the Feynman rules for Glauber operators involving soft gluon fields. We first note that the quark and gluon gauge invariant building blocks in \eq{buildingBlocks} and the mid-rapidity operator in \eq{MidRapidityOp} involve Wilson lines, and hence can source arbitrary number of gluons.
Thus, the first step in deriving the Feynman rules is to start with Wilson lines.
The momentum space expression of the soft Wilson line defined in \eq{WilsonLines} is given by
\begin{align}\label{eq:WilsonLinesMom}
\mb S_{n_i}
&= \sum_k \sum_{\rm perm} \frac{g^k}{k!}
\frac{n_i \cdot A_{s}^{A_k}(q_k) \mb T_i^{A_k}\,
n_i \cdot A_s^{A_{k-1}}(q_{k-1}) \mb T_i^{A_{k-1}} \,
\ldots
n_i \cdot A_s^{A_1} (q_{1}) \mb T_i^{A_1}}{\big [-n_i \cdot (\sum_{i = 1}^k q_i) + \im 0\big]
\big [-n_i \cdot (\sum_{i = 1}^{k-1} q_i) + \im 0\big]
\ldots
\big [-n_i \cdot q_1 + \im 0\big]}
\, ,
\end{align}
where all the momenta $q_i$ are incoming. As we mentioned above, when including Glauber operators (and Glauber bin subtractions in the soft graphs), the direction of Wilson lines becomes irrelevant and only impacts the sign of $\im0$s. In the following, whenever necessary, we will simply stick to the outgoing direction with the above prescription for $\im0$.

Next, the soft building block is defined by the relation
\begin{align}\label{eq:BSdef}
{\cal B}_{S\perp}^{A n_i \mu} &= \frac{1}{n_i\cdot {\cal P}}
n_{i\nu} \im G_S^{B\nu \mu_{\perp}} {\cal S}_n^{BA} \, ,
\qquad
{\cal S}_n^{BA} \equiv [ B | \mb S_{n_i}^{\mb g} | A ] \, ,
\end{align}
where $G_S^B$ is simply the field strength for soft gluons and ${\cal S}_{n_i}^{BA}$ is the adjoint Wilson line matrix element. From \eq{WilsonLinesMom} up to two gluons we have
\begin{align}\label{eq:SnGSExplicit}
\mb S_{n_i}^{\mb g} &=1 - g \mb T_{g}^C \frac{n_i \cdot A_{S,k}^{C}}{n_i \cdot k }
+ g^2 \bigg[
\frac{ \mb T^{C_2}_g \cdot \mb T^{C_1}_g }{n_i \cdot (k_1 + k_2) n_i \cdot k_1 }
+\frac{\mb T^{C_1}_g \cdot \mb T^{C_2}_g}{n_i \cdot (k_1 + k_2)n_i \cdot k_2 }
\bigg] \frac{n_i \cdot A_{S,k_1}^{C_1} n_i \cdot A_{S,k_2}^{C_2} }{2!} \, , \nn \\
\im G_S^{A \mu\nu} &= \Big(k^\mu g^{\nu\sigma} - k^\nu g^{\mu\sigma}\Big) A^A_{S\sigma,k }
- g \im f^{AC_1 C_2} A_{S,k_1}^{\mu C_1} A_{S,k_2}^{\nu C_2} \, ,
\end{align}
where $[B | \mb T_g^{C_1} \cdot \mb T_{g}^{C_2}| A] = \im f^{BC_1 E} \im f^{E C_2 A}$ etc. and the subscripts $k_i$ denote the incoming momenta of the gluons. We also note that matrix elements of conjugate Wilson lines $\mb S_{n_i}^{\mb g\dagger}$ simply involve transpose of the $\mb S_{n_i}^{\mb g} $ matrix in the color octet space:
\begin{align}\label{eq:SngDaggerDef}
[A | \mb S_{n_i}^{\mb g\dagger} |B ] = [B |\mb S_{n_i}^{\mb g} | A] \, .
\end{align}

To proceed further, let us write the two terms in \eq{SnGSExplicit} as
\begin{align}
\mb S_{n_i}^{\mb g} &= 1 + g \mb S_{n_i}^{\mb g[1]}(k,\mu, C) A_{S,k\mu}^C + g^2 \mb S_{n_i}^{\mb g[2]}\big ( \{k_1,\mu_1, C_1 \}, \{k_2,\mu_2, C_2 \}\big) \frac{ A_{S,k_1\mu_1}^{C_1} A_{S,k_2\mu_2}^{C_2} }{2!} \, , \\
\im G_S^{A \mu\nu} &= \im G_S^{A\mu\nu[1]} (k,\sigma, C) A_{S,k\sigma}^C + g \im G_S^{A\mu\nu[2]} \big ( \{k_1,\mu_1, C_1 \}, \{k_2,\mu_2, C_2 \}\big) \frac{ A_{S,k_1\mu_1}^{C_1} A_{S,k_2\mu_2}^{C_2} }{2!} \, .
\nn
\end{align}
These will form our building blocks for Feynman rules. For example, from \eq{BSdef} we have
\begin{align}
{\cal B}_{S\perp}^{An_i\mu} &= \frac{n_{i\nu} \im G_S^{A\nu \mu_{\perp}[1]} (k,\sigma, C) }{n_i \cdot k} A_{S,k\sigma}^C -g\Bigg[
\frac{ \im G^{A\mu\nu[2]} \big ( \{k_1,\mu_1, C_1 \}, \{k_2,\mu_2, C_2 \}\big) }{n_i\cdot (k_1 + k_2)}
\\
&\qquad + \frac{1}{n_i \cdot (k_1 + k_2)}\Big( n_{i\nu} \im G_S^{A\nu \mu_{\perp}[1]}(k_1, \mu_1 , C_1 ) \mb S_{n_i}^{\mb g[1]}(k_2,\mu_2, C_2) + (1 \lra 2)\Big)\Bigg] \frac{ A_{S,k_1\mu_1}^{C_1} A_{S,k_2\mu_2}^{C_2} }{2!} \, .
\nn \\
&\equiv
{\cal B}_{S\perp}^{An_i\mu[1]} (k,\sigma, C) A_{S,k\sigma}^C + {\cal B}_{S\perp}^{An_i\mu[2]} \big ( \{k_1,\mu_1, C_1 \}, \{k_2,\mu_2, C_2 \}\big) \frac{ A_{S,k_1\mu_1}^{C_1} A_{S,k_2\mu_2}^{C_2} }{2!} \, .
\nn
\end{align}
Thus, the two terms in the last line correspond to the Feynman rule for one and two gluon emission from $ {\cal B}_{S\perp}^{An_i\mu}$. Simplifying the first term we find
\begin{align}
{\cal B}_{S\perp}^{An_i\mu[1]} (k,\sigma, C) = \delta^{AC} \Bigg(g_\perp^{\mu \sigma} - \frac{k_\perp^\mu n_i^\sigma}{n_i\cdot k}\Bigg) \, ,
\end{align}
which we used above in \eq{BnperpExp}.

In the next step, we now derive the Feynman rule for soft gluon bilinear in \eq{SoftBilinears}:
\begin{align}
\mb O_s^{n_i g D} &= 8\pi \alpha_s A_{S,k_1\mu_1}^{C_1}A_{S,k_2\mu_2}^{C_2} \\
&\quad \times\bigg[ \frac{1}{2}\im f^{ADB} {\cal B}_{S\perp\mu}^{Bn_i[1]} (k_1,\mu_1 , C_1) \frac{n_i \cdot(k_2 - k_1)}{2} {\cal B}_{S\perp}^{An_i\mu[1]} (k_2,\mu_2 , C_2) + (1 \lra 2)\bigg] \, .\nn
\end{align}
The factor of $\im f^{ADB}$ results from the action of $\mb T_g^D$ as shown in \eq{TiAction}. We see that the two terms are in fact equal, and find the Feynman rule,
\begin{align}\label{eq:sGs}
\mb O_s^{n_i g D[1\ra1]}
&=8\pi \alpha_s \im f^{C_1DC_2} \frac{n_i \cdot (k_1 - k_2)}{2} g_{\perp \mu\nu}
\Bigg(g^{\mu\mu_1} - \frac{k_{1}^\mu n_i^{\mu_1}}{n_i\cdot k_1}\Bigg)
\Bigg(g^{\nu \mu_2} - \frac{k_{2}^\nu n_i^{\mu_2}}{n_i\cdot k_2}\Bigg) \, .
\end{align}
From \eq{OnsFull} we see that the forward scattering process only involves transfer of $\bn_i \cdot k_s$ component between the soft and collinear sector, such that the $n_i \cdot k_s$ momentum is conserved. With both $k_1$ and $k_2$ incoming in the expression above we have $n_i \cdot k_2 = -n_i \cdot k_1$.

We now turn to the mid-rapidity operator in \eq{MidRapidityOp2}. By now we have assembled all the ingredients to write down the result. We write the expansion in the analogous way:
\begin{align}\label{eq:OnsnExp}
\hat {\mb O}_s^{(n_i n_j)} (-q_\perp, q_\perp^\prime ) &= (2\pi)^2 \delta^2 (q_\perp -q_\perp') 8\pi \alpha_s q_\perp^2 \\
&\quad + (2\pi)^2 \delta^2 (q_\perp -q_\perp' - k_\perp) \hat {\mb O}_s^{(n_i n_j)[1]} (-q_\perp, q_\perp^\prime, k, \mu, C) A_{S,\mu,k}^C + \ldots \,, \nn
\end{align}
where the first term corresponds to a Glauber exchange without any soft emission.
The single gluon emission case corresponds to the Lipatov Vertex and is given by
\begin{align}
\hat {\mb O}_s^{(n_i n_j)[1]} &(-q_\perp, q_\perp^\prime, k, \mu, C) =8\pi \alpha_s g \Bigg[ q_\perp \cdot q_\perp' \big( \mb S_{n_i}^{\mb g\dagger [1]}(k,\mu, C) + \mb S_{n_i}^{\mb g [1]}(k,\mu, C) \big) \\
& + \mb T_g^{A} \bigg( q_{\perp\nu} {\cal B}_{S\perp}^{An_i\nu[1]} (k,\mu, C) + q_{\perp\nu}^\prime {\cal B}_{S\perp}^{An_j\nu[1]} (k,\mu, C) + \frac{n_{i\rho} n_{j\sigma}}{2} \im G_S^{A\rho\sigma[1]} (k,\mu, C) \bigg) \Bigg] A_{S,\mu,k}^C\nn \, .
\end{align}
Because of \eq{SngDaggerDef} the term proportional to $q_\perp \cdot q_\perp'$ vanishes and the remaining terms simplify to the following expression:
\begin{align}\label{eq:LipatovFeynRule}
\hat {\mb O}_s^{(n_i n_j)[1]} &= 8\pi \alpha_s g \mb T_g^{C} \bigg[q_\perp^\mu + q_\perp^{\prime\mu}
-\Big( \frac{n_i^\mu q_\perp^{\,2}}{n_i \cdot k}
- \frac{n_j^\mu q_\perp^{\,\prime2}}{n_j\cdot k} \Big)
- \frac{n_i\cdot k n_j \cdot k}{2}
\Big(\frac{n_i^\mu}{n_i\cdot k} - \frac{n_j^\mu}{n_j \cdot k}\Big)
\bigg]
\nn \\
&= 8\pi \alpha_s g \mb T_g^{C} \bigg[q_\perp^\mu + q_\perp^{\prime\mu}
+\Big( \frac{n_i^\mu q_\perp^{\,2}}{n_i \cdot q'} + \frac{n_j^\mu q_\perp^{\,\prime2}}{n_j\cdot q} \Big)
- \frac{n_i\cdot q' n_j \cdot q}{2}
\Big(\frac{n_i^\mu}{n_i\cdot q'} + \frac{n_j^\mu}{n_j \cdot q}\Big)
\bigg] \, .
\end{align}
This effective vertex is a combination of the three gluon emission rule and the soft emissions off the hard partons above and below, and before and after the Glauber gluon exchange. In the second line we made use of the relation $k + q' - q = 0$ and dropped the subleading $n_i \cdot q$ and $n_j\cdot q'$ components.

Similarly, by consolidating ${\cal O}(g^2)$ terms we can derive the two gluon emission Feynman rule:
\begin{align}\label{eq:MidRap2FeynRule}
&\hat {\mb O}_s^{(n_i n_j)[2]} \big (-q_\perp, q_\perp^\prime, \{k_1,\mu_1, C_1 \}, \{k_2,\mu_2, C_2 \}\big) \\
&\quad= 8\pi \alpha_s g^2 \mb T_g^{C_1} \cdot \mb T_{g}^{C_2} \Bigg[ g_\perp^{\mu_1\mu_2}
+ \frac{n_i^{\mu_1}(2 q_\perp^{\prime \mu_{2}} + k_{2\perp}^{\mu_2})}{n_i\cdot k_1}
-\frac{\big(2 q_{\perp}^{\mu_1} - k_{1\perp}^{\mu_1} \big)n_j^{\mu_2}}{n_j \cdot k_2}
+ \frac{n_i^{\mu_1}n_j^{\mu_2} - n_j^{\mu_{1}} n_i^{\mu_2} }{2}
\nn
\\
&\qquad + \frac{n_i^{\mu_{1}} n_j^{\mu_2}}{n_i\cdot k_1 n_j\cdot k_2}
\Big[
q_{\perp} \cdot q_\perp^{\,\prime}
+ k_{1\perp} \cdot k_{2\perp}
+ k_{1\perp} \cdot q_{\perp}
- k_{2\perp} \cdot q_\perp^{\,\prime}
+\frac{1}{2} (n_i\cdot k_2 n_j \cdot k_2 + n_i\cdot k_1 n_j \cdot k_1)
\Big]\nn
\\
&\qquad - \frac{n_i^{\mu_{1}}n_i^{\mu_2} }{n_i\cdot k_1} \Big(\frac{ q_{\perp}^{2}}{n_i\cdot (k_1+k_2)} + \frac{n_j \cdot k_2}{2}\Big)
- \frac{n_j^{\mu_{1}} n_j^{\mu_2}}{n_j \cdot k_2} \Big( \frac{ q_\perp^{\prime 2}}{n_j \cdot (k_1+k_2)} + \frac{n_i \cdot k_1}{2}
\Big) \Bigg] + (1 \lra 2)
\, . \nn
\end{align}

\section{Results for amplitudes with Glauber exchanges}
Here we derive the results for all the diagrams discussed in the main body of the paper. Since our results will mostly involve considering only a pair of hard partons, we will adopt the notation $n = n_i$ and $\bn = n_j$, and use $n\cdot \bn = 2$.

\subsection{One loop Glauber exchange}
\label{app:1L0R}
We first evaluate the diagram in \eq{G0ij}. These integrals are divergent and hence, to regulate them, we will make use of the rapidity regulator that inserts $|2\ell_z|^{-\eta} \nu^\eta$ at each vertex. Here $\nu$ acts as a rapidity cut off and the limit $\eta \ra 0$ is taken before $\eps \ra 0$ in dimensional regularization. For illustration we take the outgoing partons $ij$ to be a $q \bar q$ pair but keep the color matrices generic:
\begin{align}
G_0^{(ij)} &\equiv
\includegraphics[height=2.5cm,valign=c]{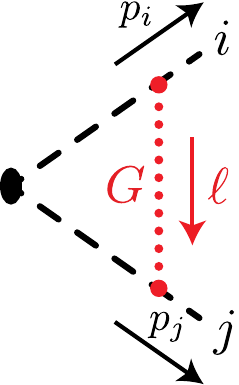}
=\im 8\pi \alpha_s \delta^{AB}\int
\bigg[\frac{ \dfbar^d \ell \:\mu^{2\eps} }{\ell_\perp^2 -m^2}\bigg]
\overline u_{n}(p_i)
\Big(\mb T_i^A\frac{\bnslash}{2}\Big)
\bigg(\frac{\nslash}{2}\frac{\im }{ P_i }\bigg)
\Gamma
\bigg(\frac{\bnslash}{2}\frac{\im }{P_j }\bigg)
\Big( \mb T_j^B\frac{\nslash}{2} \Big) v_{\bn}(p_j) \nn
\\
&= \im 8\pi \alpha_s \overline u_n(p_i) \Gamma v_{\bn}(p_j) (\mb T_i \cdot \mb T_j)
\int \frac{\df^{d-2} \ell_\perp}{(2\pi)^{d-2}}
\bigg[\frac{\mu^{2\eps} }{\ell_\perp^2 -m^2}\bigg]
\int \dfbar \ell_z \,\dfbar \ell_0 \:
\frac{(\im)^2|\ell_z|^{-2\eta} (\nu/2)^{2\eta}}{P_i P_j}\,.
\end{align}
where the color factors result from the quark operator in \eq{QuarkExplicit}. We have made use of the tree level result for $ \hat {\mb O}_s^{(n_i n_j)}$ in \eq{OnsnExp} and combined the collinear and soft operators following \eq{OnsnMom}. In the following we will drop the spinor factors and the expression will continue to hold for other cases. The propagators are given by
\begin{align}
P_i = n \cdot (p_i + \ell) + \frac{(p_{i\perp} + \ell_\perp)^2}{\bn \cdot p_i }+ \im 0 \, , \qquad
P_j = \bn \cdot (p_j- \ell) + \frac{(p_{j\perp} - \ell_\perp)^2}{n \cdot p_j } + \im 0 \,
\, .
\end{align}
Thus
\begin{align}
\int \dfbar \ell_z \int \dfbar \ell_0 \:
\frac{(\im)^2|\ell_z|^{-2\eta} (\nu/2)^{2\eta}}{P_i P_j}
&=
\int \dfbar \ell_z \int \dfbar \ell_0 \:
\frac{|\ell_z|^{-2\eta} (\nu/2)^{2\eta}}{n \cdot \ell + \Delta (\ell_\perp) + \im 0} \frac{1}{\bn \cdot \ell -\overline\Delta (\ell_\perp) - \im 0} \, .
\end{align}
where
\begin{align}\label{eq:DeltaDef}
\Delta (\ell_\perp ) = n \cdot p_i - \frac{(\vec p_{i\perp} + \vec \ell_\perp)^2}{\bn \cdot p_i} \, , \qquad
\overline \Delta (\ell_\perp ) = \bn \cdot p_j - \frac{(\vec p_{j\perp} - \vec \ell_\perp)^2}{n\cdot p_j} \, .
\end{align}
To further simplify, we first do the $\ell^0$ integral:
\begin{align}
\int \frac{\df \ell^0}{2\pi}\: \Big(\frac{1}{\ell^0 - \ell_z+ \Delta (\ell_\perp) + \im 0}\Big)\Big( \frac{1}{\ell^0 + \ell_z -\overline\Delta (\ell_\perp) - \im 0}\Big)
= \frac{\im}{2\ell_z -(\Delta(\ell_\perp) +\overline \Delta(\ell_\perp) ) - \im 0}
\, ,
\end{align}
such that~\cite{Rothstein:2016bsq}
\begin{align}
\int \dfbar \ell_z \int \dfbar \ell_0 \:
\frac{|\ell_z|^{-2\eta} (\nu/2)^{2\eta}}{P_i P_j}=
\int \dfbar \ell_z \frac{\im|\ell_z|^{-2\eta} (\nu/2)^{2\eta}}{2\ell_z -(\Delta(\ell_\perp) +\overline \Delta(\ell_\perp) ) - \im 0} = \frac{1}{4}
\, ,
\end{align}
Hence,
\begin{align}
G_0^{(ij)} &=
\im 2\pi \alpha_s (\mb T_i \cdot \mb T_j)
\int \frac{\df^{d-2} \ell_\perp}{(2\pi)^{d-2}}
\bigg[\frac{\mu^{2\eps} }{\ell_\perp^2 -m^2}\bigg] = \mb C^{(ij)} (m, \mu) \, .
\end{align}
where $\mb C^{(ij)}(m,\mu)$ was stated in \eq{CijmuDef}.

\subsection{Single soft emission graphs}

\subsubsection{Lipatov vertex graph}
\label{app:Lipatov}
Now we turn to the graph (b) in \fig{1L1R}. This graph contains the Lipatov vertex whose Feynman rule we derived above in \eq{LipatovFeynRule}.
For simplicity, we will set $q_{1\perp} = q_{1\perp}^{(ij)}$ in \fig{1L1R}b.
Using this Feynman rule and carrying out integration over the longitudinal momenta as above, we have
\begin{align}\label{eq:G1bExplicit}
G_{1(b)}^{(ij)} &\equiv \includegraphics[height=2.5cm,valign=c]{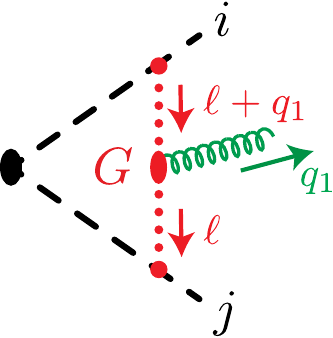}
= \im g8 \pi \alpha_s\big[\mb T_i^A (\im f^{AC_1B}) \mb T_j^B\big] \frac{1}{4}\int \frac{\mu^{2\eps} \:\dfbar^{d-2} \ell_\perp}{(\ell_\perp^2 -m^2) \,[ (\ell_\perp+ q_{1\perp})^2 -m^2]}
\\&\quad \times\veps_{ \mu}^{*}(q_1)
\bigg[
- (2\ell_\perp^\mu + q_{1\perp}^\mu)
+ \Big(\frac{n^\mu}{n\cdot q_1} (\ell_\perp + q_{1\perp})^2
- \frac{\bn^\mu}{\bn \cdot q_1} \ell_\perp^2
\Big)
- \frac{q_{1\perp}^2}{2} \Big(\frac{n^\mu}{n\cdot q_1} - \frac{\bn^\mu}{\bn \cdot q_1}\Big)
\bigg] \, . \nn
\end{align}
The first term in the second line integrates to zero, as can be seen by first changing the variables $\ell_\perp+ \frac{q_{1\perp}}{2} = k_\perp$ and noting that any term proportional to $k_\perp^\mu$ in the numerator vanishes. Thus,
\begin{align}
G_{1(b)}^{(ij)} = \im g 2\pi \alpha_s \big[(\mb T_i \cdot \mb T_j ), \mb d^{(0)C_1}_{ji}(q_1)\big]
\bigg(\frac{q_{1\perp}^2}{2} \int \frac{\mu^{2\eps} \: \dfbar^{d-2} \ell_\perp}{(\ell_\perp^2 -m^2) \,[ (\ell_\perp+ q_{1\perp})^2 -m^2]} - \int \frac{\mu^{2\eps} \: \dfbar^{d-2} \ell_\perp}{(\ell_\perp^2 -m^2)} \bigg) \, .
\end{align}
Using the defining equation for $\mb C^{(ij)} (a,b)$ in \eq{CijDef} and the results in \eqs{perpInteg}{ThetaFunc} we find
\begin{align}\label{eq:Cmq1}
\im g 2\pi \alpha_s \mb T_i \cdot \mb T_j \frac{q_{1\perp}^2}{2} \int \frac{\mu^{2\eps} \: \dfbar^{d-2} \ell_\perp}{(\ell_\perp^2 -m^2) \,[ (\ell_\perp+ q_{1\perp})^2 -m^2]} = \mb C^{(ij)} (m, q_{1\perp}^{(ij)}) \, .
\end{align}
Combining this wih the third term in \eq{G1bExplicit} we arrive at the result in \eq{G1b0}:
\begin{align}\label{eq:G1bij}
G_{1(b)}^{(ij)} &= g
\big[\mb C^{(ij)} (m, q_{1\perp}^{(ij)}) - \mb C^{(ij)} (m, \mu) , \mb d^{(0)C_1}_{ji}(q_1)\big] \nn \\
&= g [C_1|\big[ \mb d^{(0)}_{ji} (q_1), \: \mb C^{(ij)}(q_{1\perp}^{(ij)}, \mu)\big ]
\, .
\end{align}
\subsubsection{Soft gluon rescattering graphs}
\label{app:Rescatter}
Next we consider the graph \fig{1L1R}c that corresponds to rescattering between the soft gluon and one of the hard partons. This involves T-product of $O_{ns}^{ij}$ operator insertion in \eq{OnsFull}, using the Feynman rule for $sGs$ vertex in \eq{OnsFull}, and emission via the $\mb S_{n_i}$ Wilson line, using \eq{WilsonLinesMom}:
\begin{align}\label{eq:G1cnaive}
&\widetilde {G}_{1(c)}^{(ij)} \equiv \includegraphics[height=2.5cm,valign=c]{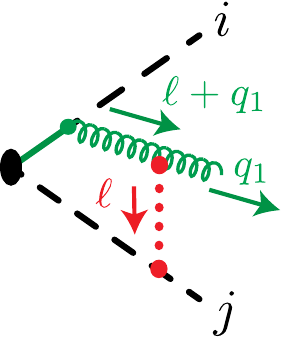}
= \im \veps_{\sigma}^* (q_1)
\int \frac{\mu^{2\eps} \:\dfbar^{d} \ell\, |\ell_z|^{-2\eta}(\nu/2)^{2\eta}}{(\ell_\perp^2 -m^2)}
\frac{g \mb T_i^C n_\rho }{n\cdot (q_1 + \ell) + \im 0}
\frac{-\im}{(q_1 + \ell)^2 -m^2+ \im 0} \nn \\
&\quad \times 8\pi\alpha_s \im f^{CDC_1} \bn \cdot q_1 g_{\perp \mu\nu} \Bigg(g^{\mu\rho} - \frac{(\ell^\mu+ q_1^\mu)\bn^{\rho}}{\bn \cdot q_1} \Bigg)\Bigg(g^{\nu\sigma} - \frac{q_{1}^\nu \bn^\sigma}{\bn \cdot q_1}\Bigg) \frac{-\im \mb T_j^D }{\big[\bn \cdot \ell - \overline \Delta (\ell_\perp) - \im 0\big]}
\end{align}
The loop momentum component $\bn \cdot \ell \sim \lambda^2$ in \fig{1L1R}c, which is why we dropped it in comparison with $\bn \cdot q_1 \sim \lambda$. The denominator also simplifies:
\begin{align}
(q_1+\ell)^2 -m^2 + \im 0 = \bn \cdot q_1 (n \cdot \ell + \Delta_{q_1}(\ell_\perp) + \im 0 ) \, , \quad
\Delta_{q_1}(\ell) = n\cdot q_1 - \frac{(\vec q_{1\perp} + \vec \ell_\perp)^2 + m^2}{\bn \cdot q_1} \, .
\end{align}
Hence we see that the naive graph vanishes on account of there being two poles on both the sides. Hence, the result is actually given by negative of the zero-bin graph where the combination $n\cdot (q_1+\ell) \sim \lambda^2$ instead of ${\cal O}(\lambda)$ in \eq{G1cnaive}:
\begin{align}
G_{1(c)}^{(ij)} &= \im 16 \pi \alpha_s g \mb T_i^C \im f^{CDC_1} \mb T_j^D
\int \frac{\mu^{2\eps} \:\dfbar^{d-2} \ell_\perp}{(\ell_\perp^2 -m^2)}
\Big(\veps^*(q_1) - q_{1} \frac{\bn \cdot \veps^*(q_1)}{\bn \cdot q_1}\Big)
\cdot
\frac{ (\ell_\perp+ q_{1\perp})}{[(\ell_\perp+ q_{1\perp})^2 - m^2]}
\\
&\times
\int \dfbar \ell_z \: \dfbar \ell_0 \: \frac{1}{n\cdot (q_1 + \ell)+ \im 0} \frac{(-\im )^2|\ell_z|^{-2\eta}(\nu/2)^{2\eta}}{\big[\bn \cdot \ell - \overline \Delta_{p_j}(\ell_\perp) - \im 0\big]} \nn \\
&=- \im 4 \pi \alpha_s g \mb T_i^C \im f^{CDC_1} \mb T_j^D
\Big(\veps^*(q_1) - q_{1} \frac{\bn \cdot \veps^*(q_1)}{\bn \cdot q_1}\Big) \cdot q_{1\perp}
\int \frac{\mu^{2\eps} \:\dfbar^{d-2} \ell_\perp}{(\ell_\perp^2 -m^2)}\frac{1}{[(\ell_\perp+ q_{1\perp})^2 - m^2]} \nn
\end{align}
where the term proportional to $\ell^\mu_{\perp}$ integrates to zero. Simplifying
$$
\mb T_i^C \im f^{CDC_1} \mb T_j^D =
- [ C_1 | \mb T_{q_1}^D \mb T_j^D | C ][ C | \mb T_i
= - [ C_1 | \big (\mb T_{q_1} \cdot \mb T_j \big) \mb T_i \, ,
$$
and
$$
\Big(\veps^*(q_1) - q_{1} \frac{\bn \cdot \veps^*(q_1)}{\bn \cdot q_1}\Big) \cdot q_{1\perp} = \frac{q_{1\perp}^2}{2} \Big(\frac{n\cdot \veps^*(q_1)}{n\cdot q_1} - \frac{\bn \cdot \veps^*(q_1)}{\bn\cdot q_1}\Big) \, ,
$$
and using result in \eq{Cmq1}, we arrive at
\begin{align}\label{eq:G1cRes}
G_{1(c)}^{(ij)} &=g \: [C_1 | \mb C^{(q_1 j)} (m, q_{1\perp}^{(ij)}) \: \mb d^{(0)}_{ji}(q_1) \, .
\end{align}
Finally, graphs where the Glauber exchanges happen with the same parton that sourced the soft gluon vanish in Feynman gauge.

\subsection{Double soft emission graphs}

\subsubsection{Mid-rapidity double soft emission graphs}
\label{app:midRapidity}
Here we derive results for the double soft emission graphs from $i$-$j$ forward scattering sub-process shown in \fig{2gall}e-h. The graph \fig{2gall}e is a combination of Lipatov vertex and 3-gluon vertex, and is given by
\begin{align}
&S_{2(e)}^{(ij)}
\equiv \includegraphics[height=2.5cm,valign=c]{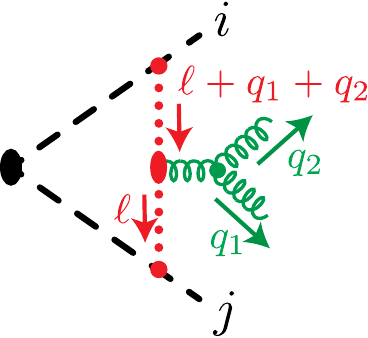}
\\
&=
g \veps_1^{\mu_1} \veps_2^{\mu_2}
\big[ \mb d_{ji}^{ \mu (0)A} (q_1+q_2), \mb C^{(ij)} (q_{1\perp}^{(ij)} + q_{2\perp}^{(ij)}, \mu ) \big]
\Big(\frac{-\im}{(q_1 +q_2)^2}\Big)
\Big[-g V_{\mu \mu_1\mu_2}^{A C_1 C_2}(q_1+q_2, -q_1, -q_2)\Big]
\nn \\
&= g^2 \big[ \mb d_{ji}^{ \mu (0)A} (q_1+q_2), \mb C^{(ij)} (q_{1\perp}^{(ij)} + q_{2\perp}^{(ij)}, \mu ) \big]
\frac{\im g f^{AC_1C_2}\big[
\veps_1 \cdot \veps_2 (q_{2\mu} - q_{1\mu})
- \veps_{2\mu} (2 q_2 \cdot \veps_1)
+ \veps_{1\mu} (2 q_1 \cdot \veps_2)
\big]}{(q_1+q_2)^2} \, .\nn
\end{align}
In the ordered limit $q_2^\mu\ll q_1^\mu$, we find
\begin{align}
S_{2(e)}^{(ij)} &= g^2 \frac{\im g f^{C_1 C_2A}\veps_2 \cdot q_1}{q_2 \cdot q_1} \big[ \mb d_{ji}^{ \mu (0)A} (q_1), \mb C^{(ij)} (q_{1\perp}^{(ij)}, \mu ) \big] \, ,
\end{align}
such that, upon adding other vanishing commutators as in \eq{G1b}, the result reduces to \eq{G2ef}.

Next, we consider the 2 soft gluon emissions from the mid-rapidity operator in \fig{2gall}f. Using the Feynman rule in \eq{MidRap2FeynRule} and performing longitudinal momenta integration we find
\begin{align}\label{eq:S2f}
S_{2(f)}^{(ij)} \equiv \includegraphics[height=2.5cm,valign=c]{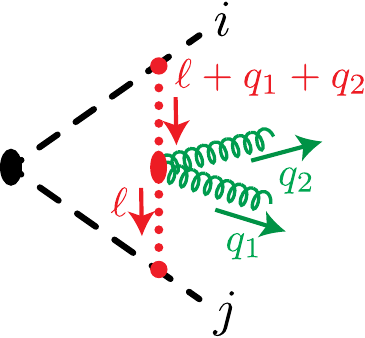}
&=\im \veps^{*\mu_1}(q_1)\veps^{*\mu_2}(q_2) \frac{1}{4} \int \frac{\dfbar^{d-2} \ell_\perp\: \mu^{2\eps}\, \mb T_i^A [A |\hat {\mb O}_{s\mu_1\mu_2}^{(n_i n_j)[2]} |B] \mb T_i^B}{[(\ell_\perp + q_{1\perp}+q_{2\perp})^2 -m^2 ](\ell_\perp^2-m^2)} \, ,
\end{align}
where the momentum flow is as shown above. Now, let $\ell_{t\perp} = \ell_\perp + q_{1\perp} + q_{2\perp}$. In order to cancel as many terms as possible in the integration we re-express $\ell_\perp$ as
\begin{align}
\ell_\perp = k_\perp - \frac{q_{1\perp} + q_{2\perp}}{2} \, ,
\qquad
\ell_{t\perp} = k_\perp + \frac{q_{1\perp} + q_{2\perp}}{2} \, ,
\end{align}
as a result of which terms proportional to $\vec k_\perp$ will integrate to zero after azimuthal integration. The remaining terms are given by
\begin{align}\label{eq:Os2simpler}
\int \frac{ d\phi }{2\pi}\hat {\mb O}_{s}^{(n_i n_j)[2]\mu_1\mu_2} &
= 8\pi \alpha_s g^2\, \mb T_g^{C_1} \cdot \mb T_g^{C_2} \Bigg[
g_\perp^{\mu_1\mu_2}
- \frac{n^{\mu_1}q_{1\perp}^{\mu_2}}{ q_1^+}
- \frac{q_{2\perp}^{\mu_1}\bn^{\mu_2}}{ q_2^-}
\\
& + \frac{n^{\mu_1}\bn^{\mu_2} - \bn^{\mu_{1}} n^{\mu_2}}{2}
+ \frac{n^{\mu_{1}} \bn^{\mu_2}}{2 q_1^+ q_2^-}
\Big[ \ell_\perp^{ 2} + \ell_{t\perp}^{ 2}
- ( q_{1\perp} - q_{2\perp})^2
\Big]\nn
\\
&
- \frac{n^{\mu_{1}}}{q_1^+} \frac{n^{\mu_2}}{q_2^+} \Big( \ell_{t\perp}^{\,2}\frac{ q_2^+}{ (q_1^++q_2^+)} - \frac{ q_{2\perp}^{2}}{2}\Big)
- \frac{\bn^{\mu_{1}}}{q_1^-} \frac{\bn^{\mu_2}}{q_2^-} \Big(
\ell_\perp^{\,2}\frac{q_1^-}{(q_1^-+q_2^-)} - \frac{ q_{1\perp}^{\,2}}{2}
\Big) \Bigg] + (1 \lra 2)
\, .\nn
\end{align}
We now simplify the expression above further by integrating over $\ell_\perp$ and further dropping terms that are subleading in the limit $q_2^{\mu} \ll q_1^\mu$. This results in
\begin{align}
S_{2(f)}^{(ij)} &= \im 2\pi \alpha_s g^2 \mb T_i^A \Big[ A \Big | \bigg ( \frac{\bn \cdot \veps^*(q_2)}{\bn \cdot q_2} \mb T_g^{C_1} \cdot \mb T_g^{C_2} - \frac{n\cdot \veps^*(q_2)}{n\cdot q_2} \mb T_g^{C_2} \cdot \mb T_g^{C_1}\bigg) \Big | B \Big ] \mb T_j^B
\\
& \quad\times \Big(\frac{n \cdot \veps^*(q_1) }{n\cdot q_1} - \frac{\bn \cdot \veps^*(q_1) }{\bn \cdot q_1}\Big)
\bigg(\frac{q_{1\perp}^2}{2} \int \frac{\mu^{2\eps} \: \dfbar^{d-2} \ell_\perp}{(\ell_\perp^2 -m^2) \,[ (\ell_\perp+ q_{1\perp})^2 -m^2]} - \int \frac{\mu^{2\eps} \: \dfbar^{d-2} \ell_\perp}{(\ell_\perp^2 -m^2)} \bigg) \, .\nn
\end{align}
Next we simplify the color factors so as to sequentially source $q_1$ and $q_2$ gluons from $i$ and $j$:
\begin{align}
\mb T_i^A [A | \mb T_g^{C_1} \cdot \mb T_g^{C_2} |B] \mb T_j^{B} &=
-\big[\im f^{C_1 A B} \mb T_i^A \mb T_j^B , \mb T_j^{C_2} \big]
=- \Big[ \big[\mb T_i \cdot \mb T_j , \, \mb T_i^{C_1}\big], \, \mb T_j^{C_2}
\Big]
\\
\mb T_i^A [A | \mb T_g^{C_2} \cdot \mb T_g^{C_1} |B] \mb T_j^{B}
&=
\big[\im f^{C_1 A B} \mb T_i^A \mb T_j^B , \, \mb T_i^{C_2} \big]
= \Big[ \big[\mb T_i \cdot \mb T_j , \, \mb T_i^{C_1}\big], \, \mb T_i^{C_2}
\Big] \, , \nn
\end{align}
which yields
\begin{align}
S_{2(f)}^{(ij)} &=g^2 \Bigg[ \Big(\mb T_i^{C_2} \frac{n\cdot \veps^*(q_2)}{n\cdot q_2} +\mb T_j^{C_2} \frac{\bn\cdot \veps^*(q_2)}{\bn\cdot q_2} \Big) , \Big[ \mb d^{(0)}_{ji}(q_1) , \, \mb C^{(ij)}(q_{1\perp}^{(ij)}, \mu)\Big] \Bigg] \nn \\
&= g^2\Big[ \tilde{\mb J}^{(0)C_2}(q_{2}) , \, \big[
\tilde {\mb J}^{(0)C_1}(q_1) , \, \mb C^{(ij)} (q_{1\perp}^{(ij)}, \mu)\big]\Big] \, ,
\end{align}
where in the last line we have added the remaining vanishing commutators to simplify the result.

\subsubsection{Soft propagator graphs}
\label{app:SoftProp}
The graphs (g) and (h) in \fig{2gall} involve a T-product of $O_{n_is}$ and $O_{n_js}$ insertions and a soft propagator. We will find that these graphs are subleading in the ordered limit.
The momenta of the propagators from top to bottom for diagram (g) are assigned as follows:
\begin{align}
\ell_1^\mu &= \ell_\perp^\mu + q_{1\perp}^\mu + \bn \cdot (q_1 + q_2) \frac{n^\mu}{2} + n \cdot \ell \frac{\bar n^\mu}{2}\\
\ell_2^\mu &= \ell_\perp^\mu - n\cdot q_1 \frac{\bn^\mu}{2} + \bn \cdot q_2 \frac{n^\mu}{2} \nn \\
\ell_3^\mu &= \ell_\perp^\mu - q_{2\perp}^\mu - n \cdot (q_1 + q_2) \frac{\bn^\mu}{2} + \bar n \cdot \ell \frac{n^\mu}{2} \nn
\end{align}
where the loop momentum $\ell^\mu$ scales as
\begin{align}
\ell^\mu \sim (\lambda^2, \lambda^2 ,\lambda) \, .
\end{align}
The $n$ and $\bn$ collinear quark propagators involve
\begin{align}
P_i &= n \cdot (p_i - \ell) - \frac{(\vec p_{i\perp} - \vec \ell_\perp -\vec q_{1\perp})^2}{\bn \cdot p_i} + \im 0 \,, \nn \\
P_j &= \bn \cdot (p_j + \ell) - \frac{(\vec p_{j\perp} + \ell_\perp - \vec q_{2\perp})^2}{n \cdot p_j} + \im 0 \, ,
\end{align}
such that the $\ell^\pm$ momenta can be integrated over as before.
Then, the graph (g) using the Feynman rule in \eq{sGs} and integrating over the longitudinal momenta is given by
\begin{align}\label{eq:SoftPropDrop}
S_{2(g)}^{(ij)} \equiv \includegraphics[height=2.5cm,valign=c]{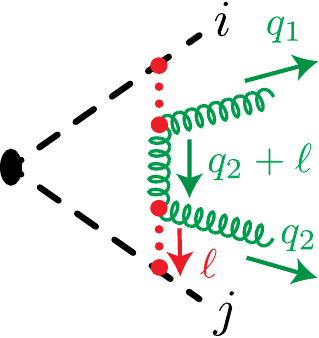}
&= \im 4\pi \alpha_s g^2 \mb T_i^A \mb T_j^B (\im f^{AC_1E} )(\im f^{BC_2E}) \Big(\veps_1^\alpha - q_{1}^\alpha \frac{n\cdot \veps_1}{n\cdot q_1}\Big) \Big(\veps_2^\beta - q_{2}^\beta \frac{\bn\cdot \veps_2}{\bn\cdot q_2}\Big) \nn \\
&\times \int \dfbar^2 \ell_\perp
\frac{ \big(n\cdot q_1 \bn \cdot q_2 \, g_\perp^{\alpha\beta} + 2 \ell_\perp^\alpha \ell_\perp^\beta\big) }{(\ell_\perp + q_{1\perp}) ^2(\ell_\perp - q_{2\perp})^2(-n\cdot q_1 \bn \cdot q_2 + \ell_\perp^2 + \im 0 )} \, .
\end{align}
To see why this graph is subleading in the limit $q_{2\perp} \ll q_{1\perp}$ we first note that the transverse momentum injected from $p_i$ must be at least $\ell_\perp \sim q_{1\perp}$. This implies that the integral in the second line is roughly of the order $q_{1\perp}^\alpha q_{1\perp}^\beta/q_{1\perp}^4$, which when dotted with the vectors in the first line does not lead to terms that scale as $1/n_{i,j}\cdot q_2$, unlike every other leading graph.

\subsubsection{Soft gluon rescattering graphs}
\label{app:rescatter2}
We now consider the graphs shown in \fig{2gall}i-l that involve double soft emission and Glauber exchange between the softer emission $q_2$ and a collinear leg. The graph (i) is given by
\begin{align}
&\widetilde {S}_{2(i)}^{(ij),(q_2j)} \equiv
\includegraphics[height=2.5cm,valign=c]{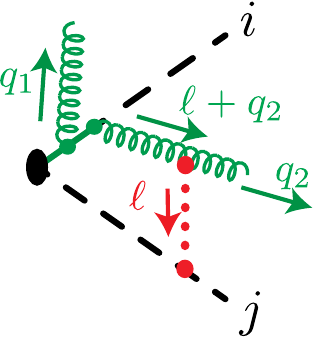}
= \im 16 \pi \alpha_s g^2
\int \frac{\mu^{2\eps} \:\dfbar^{d-2} \ell_\perp}{(\ell_\perp^2 -m^2)}
\Big(\veps_2 - q_{2} \frac{n_j \cdot \veps_2}{n_j \cdot q_2}\Big) \cdot(\ell_\perp+ q_{2\perp}) ( \im f^{ABC_2} ) \nn \\
&\times \int \dfbar \ell_z\, \dfbar \ell_0 \:
\frac{n_i \cdot \veps_1\,\mb T_i^{C_1} }{n_i\cdot(q_1 + q_2 + \ell) + \im 0 } \frac{\mb T_i^B}{n_i\cdot (q_2 + \ell) + \im 0}
\frac{-\im}{(q_2 + \ell)^2 -m^2+ \im 0} \frac{-\im |\ell_z|^{-2\eta}(\nu/2)^{2\eta} \mb T_j^A}{\big[n_j \cdot \ell - \overline \Delta (\ell_\perp) - \im 0\big]} \, .
\end{align}
As above, the naive graph evaluates to zero. We then consider the zero bin where $n_i \cdot (q_2 + \ell )\sim \lambda^2$, such that
\begin{align}
{S}_{2(i)}^{(ij),(q_2j)}
&= g^2\: [C_1, C_2 | \mb C^{(q_2 j)} (m, q_{2\perp}^{(ij)})\: \mb d^{(0)}_{ji}(q_2)\: \frac{\mb T_i\, n_i\cdot \veps_1}{n_i \cdot q_1} \, .
\end{align}
The graph \fig{2gall}j vanishes as the zero bin graph does not exist in this case. Adding the graph \fig{2gall}k and ones involving attachments to other collinear legs, we get
\begin{align}\label{eq:Sq2jFull}
{S}_{2(i+k)}^{(ij),(q_2j)}
&= g^2 \: [C_1, C_2 | \mb C^{(q_2 j)} (m, q_{2\perp}^{(ij)})\: \mb d^{(0)}_{ji}(q_2)\: \mb J^{(0)} (q_1)\, .
\end{align}

Now, we turn to graph \fig{2gall}l:
\begin{align}
&\widetilde {S}_{2(l)}^{(ij),(q_2j)} \equiv
\includegraphics[height=2.5cm,valign=c]{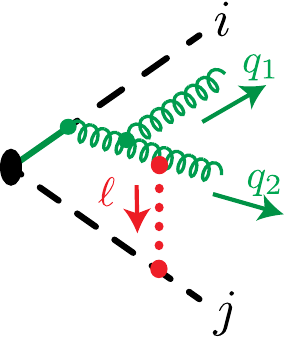}
\\
&= \im 8\pi \alpha_s g^2 \mb T_j^A( \im f^{ABC_2} ) (\im f^{BC_1 D})\mb T_i^D
\int \frac{\mu^{2\eps} \:\dfbar^{d-2} \ell_\perp}{(\ell_\perp^2 -m^2)}
g_{\perp\rho\sigma}
\Big(\veps_{2\perp}^\rho - q_{2\perp}^\rho \frac{n_j \cdot \veps_2}{n_j \cdot q_2}\Big)
\Big(g_{\perp}^{\mu\sigma}- (\ell_\perp + q_{2\perp})^\sigma \frac{n_j^\mu}{n_j\cdot q_2}
\Big)
\nn \\
&\qquad \times \int \dfbar \ell_z \,\dfbar \ell_0 \:
\frac{-\im(n_j\cdot q_2)\big[\veps_{1\mu} n_i \cdot(q_1 - q_2 - \ell) - n_i \cdot \veps_1 (2q_{1} + \ell + q_2)_\mu + 2n_{i\mu} \veps_1 \cdot (q_2 + \ell) \big]}{[n_i \cdot(q_1+q_2+\ell) + \im 0][(q_1+q_2+\ell)^2 - m^2 + \im 0]}\nn \\
&\qquad\times \frac{-\im}{(q_2 + \ell)^2 -m^2+ \im 0} \frac{-\im |\ell_z|^{-2\eta}(\nu/2)^{2\eta}}{\big[n_j \cdot \ell - \overline \Delta (\ell_\perp) - \im 0\big]} \, .
\nn
\end{align}
The zero bin corresponds to the configuration $n_i \cdot (q_1 + q_2 + \ell) \sim \lambda^2$. The choice $n_i \cdot (q_2 + \ell) \sim \lambda^2$ leads to a single pole in $n_j \cdot \ell$ and corresponds to an unphysical contribution of a Lipatov vertex between a soft and a collinear emission. This contribution is to be included separately by considering it in \eftnp where $q_1$ becomes a collinear mode. Hence, we have
\begin{align}
{S}_{2(l)}^{(ij),(q_2j)}
&= \im 2\pi \alpha_s g^2 \mb T_j^A( \im f^{ABC_2} ) (\im f^{BC_1 D})\mb T_i^D
\\
&\times
\int \frac{\mu^{2\eps} \:\dfbar^{d-2} \ell_\perp}{(\ell_\perp^2 -m^2)}
g_{\perp\rho\sigma}
\Big(\veps_{2\perp}^\rho - q_{2\perp}^\rho \frac{n_j \cdot \veps_2}{n_j \cdot q_2}\Big)
\Big(g_{\perp}^{\mu\sigma}- (\ell_\perp + q_{2\perp})^\sigma \frac{n_j^\mu}{n_j\cdot q_2}
\Big)
\nn \\
& \times
\frac{-\im(n_j\cdot q_2)\big[2 \veps_{1\mu} n_i \cdot q_1 - n_i \cdot \veps_1 (2q_{1} + \ell + q_2)_\mu + 2n_{i\mu} \veps_1 \cdot (q_2 + \ell) \big]}{[-n_i\cdot q_1 n_j\cdot q_2 + (q_{2\perp} + \ell_\perp)^2 -m^2+ \im 0][(q_{1\perp}+q_{2\perp}+\ell_\perp)^2 - m^2 + \im 0]}\nn \, .
\end{align}
This graph corresponds to the same configurations as the soft propagator graphs in \eq{SoftPropDrop} and hence is subleading in the limit $q_2^\mu \ll q_1^\mu$.

For the graphs, where the Glauber exchange happens between $q_1$ and another collinear leg we simply state the final result from \Ref{Martinez:2016vur} for these graphs (evaluated in the full theory using cut rules):
\begin{align}\label{eq:S2jq1j}
S_{2}^{(ij)(q_1j)} = g^2 \big[C_1,C_2\big|\mb C^{(q_1j)} (m, q_{1\perp}^{(ij)}) \mb J^{(0)}_2(q_2,q_1) \mb d^{(0)}_{ji}(q_1)\, .
\end{align}
This expression actually corresponds to using the tree level double soft amplitude $\mb K_2(q_1+\ell, q_2)$ in the ordered limit given by \eq{treeFactorise} within the integral, such that the result for these graphs ends up being a direct generalization of \eq{G1cRes} with an extra factor of $\mb J^{(0)}_2(q_2,q_1)$ in between. When combined with the graphs where $q_2$ is sourced by the Lipatov vertex between $q_1$ and another hard parton, given in \fig{1L2R}c (along with the relevant terms from the Wilson coefficient), we recover the expected ordering, as in the second line in \eq{AMFS_SCET}.

\bibliography{qcd}
\end{document}